\documentclass[letterpaper,twocolumn,10pt]{article}
\usepackage{usenix-2020-09}

\usepackage{url}
\usepackage{graphicx}
\usepackage{listings}
\usepackage{xcolor}
\usepackage{balance}
\usepackage{csquotes}
\usepackage[T1]{fontenc}
\usepackage{hyperref}
\usepackage{booktabs}
\usepackage{multirow}
\usepackage[most]{tcolorbox}
\usepackage{capt-of}
\usepackage{enumitem}
\usepackage{makecell, multirow, tabularx}
\usepackage[singlelinecheck=false]{caption}
\usepackage{xurl}

\definecolor{eclipseStrings}{RGB}{42,0.0,255}
\definecolor{eclipseKeywords}{RGB}{127,0,85}
\colorlet{numb}{magenta!60!black}

\lstdefinelanguage{json}{
    basicstyle=\normalfont\ttfamily,
    commentstyle=\color{eclipseStrings}, 
    stringstyle=\color{eclipseKeywords}, 
    numbers=left,
    numberstyle=\scriptsize,
    stepnumber=1,
    numbersep=8pt,
    basicstyle=\scriptsize,
    showstringspaces=false,
    breaklines=true,
    frame=lines,
    string=[s]{"}{"},
    comment=[l]{:\ "},
    morecomment=[l]{:"},
    literate=
        *{0}{{{\color{numb}0}}}{1}
         {1}{{{\color{numb}1}}}{1}
         {2}{{{\color{numb}2}}}{1}
         {3}{{{\color{numb}3}}}{1}
         {4}{{{\color{numb}4}}}{1}
         {5}{{{\color{numb}5}}}{1}
         {6}{{{\color{numb}6}}}{1}
         {7}{{{\color{numb}7}}}{1}
         {8}{{{\color{numb}8}}}{1}
         {9}{{{\color{numb}9}}}{1}
}
\hypersetup{colorlinks=true, linkcolor=blue}

\newcommand*\dash{\unskip\kern.16667em---\penalty\exhyphenpenalty
        \hskip.16667em\relax
}

\newtcolorbox[auto counter]{attack}[2][]{
  enhanced,
  fonttitle=\scshape,
  breakable,
  left=2pt,right=2pt,top=2pt,bottom=2pt,
  title={Risk~\thetcbcounter: #1},
  #2
}

\makeatletter
\newcommand{\thickhline}{%
    \noalign {\ifnum 0=`}\fi \hrule height 1pt
    \futurelet \reserved@a \@xhline
}
\newcolumntype{"}{@{\hskip\tabcolsep\vrule width 1pt\hskip\tabcolsep}}
\makeatother

\newcommand{\thickcline}[1]{%
    \@thickcline #1\@nil%
}

\def\@thickcline#1-#2\@nil{%
  \omit
  \@multicnt#1%
  \advance\@multispan\m@ne
  \ifnum\@multicnt=\@ne\@firstofone{&\omit}\fi
  \@multicnt#2%
  \advance\@multicnt-#1%
  \advance\@multispan\@ne
  \leaders\hrule\@height1pt\hfill
  \cr
  \noalign{\vskip-1pt}%
}

\newlength{\Oldarrayrulewidth}
\newcommand{\Cline}[2]{%
  \noalign{\global\setlength{\Oldarrayrulewidth}{\arrayrulewidth}}%
  \noalign{\global\setlength{\arrayrulewidth}{#1}}\cline{#2}%
  \noalign{\global\setlength{\arrayrulewidth}{\Oldarrayrulewidth}}}

\usepackage[absolute,overlay]{textpos}

\textblockorigin{0mm}{0mm}

\begin{document}

\title{\Large \bf LLM Platform Security: \\Applying a Systematic Evaluation Framework to OpenAI's ChatGPT Plugins}

\author{
\rm Umar Iqbal\\
Washington University in St. Louis
\and
\rm Tadayoshi Kohno\\
University of Washington
\and
\rm Franziska Roesner\\
University of Washington
}

\begin{textblock*}{0.95\textwidth}(0.8in,0.5in)
    \noindent 
    A shortened version of this paper appears in the \textit{7th AAAI / ACM Conference on AI, Ethics, and Society, October 2024}. This is the extended version.
\end{textblock*}


\maketitle

\thispagestyle{plain}
\pagestyle{plain}

\begin{abstract}
Large language model (LLM) platforms, such as ChatGPT, have recently begun offering an \textit{app ecosystem} to interface with third-party services on the internet.
While these apps extend the capabilities of LLM platforms, they are developed by arbitrary third parties and thus cannot be implicitly trusted. 
Apps also interface with LLM platforms and users using natural language, which can have imprecise interpretations.
In this paper, we propose a framework that lays a foundation for LLM platform designers to analyze and improve the security, privacy, and safety of current and future third-party integrated LLM platforms.
Our framework is a formulation of an attack taxonomy that is developed by iteratively exploring how LLM platform stakeholders could leverage their capabilities and responsibilities to mount attacks against each other. 
As part of our iterative process, we apply our framework in the context of OpenAI's plugin (apps) ecosystem. 
We uncover plugins that concretely demonstrate the potential for the types of issues that we outline in our attack taxonomy. 
We conclude by discussing novel challenges and by providing recommendations to improve the security, privacy, and safety of present and future LLM-based computing platforms.
\end{abstract}

\section{Introduction}
\label{sec:introduction}

Large language models (LLMs), such as GPT-4~\cite{gpt4_link}, and platforms that leverage them, such as ChatGPT~\cite{chatgpt_announcement}, have advanced tremendously in capabilities and popularity.
In addition to the actual LLM at their core, platforms like ChatGPT~\cite{chatgpt_announcement} and Bard~\cite{google_bard_url} are becoming increasingly complex in order to support various use cases and integrate with different features and third-party services. 
For example, platform vendors like OpenAI and Google have announced and begun implementing app ecosystems, allowing the LLM to interface with third-party services~\cite{openai_chatgpt_plugins,google_bard_plugin_announcement}.
In this paper, we investigate conceptually and empirically the security of these emerging LLM-based platforms that support third-party integrations.
In this paper we focus on OpenAI's plugins as a case study (Note that Plugins have now transitioned into Actions, which follow a similar structure and are embedded in OpenAI's GPTs~\cite{openai-action}).

While extending the capabilities of LLM platforms, third-party plugins may add to the long list of security, privacy, and safety concerns raised by the research community about LLMs, e.g.,~\cite{greshake2023not,jakesch2023co,kang2023exploiting,PerezPromptInjectionNeurIPSWorkshop,zou2023universal,bagdasaryan2023ab}. 
First, plugins are developed by third-party developers and thus should not be implicitly trusted. 
Prior research on other computing platforms has shown that third-party integrations often raise security and privacy issues, e.g.,~\cite{enck2011study,MayerThirdPartyTracking12,FernandesIoTAppSecSP16, farooqi2020canarytrap, ifttt, chen2022experimentalSlack}. 
In the case of LLM platforms, anecdotal evidence already suggests that third-party plugins can launch prompt injection attacks and can potentially take over LLM platforms~\cite{propmpt_injection_plugins}.
Second, as we observe, plugins interface with LLM platforms and users using natural language, which can have ambiguous and imprecise interpretation.
For example, the natural language functionality descriptions of plugins could either be interpreted too broadly or too narrowly by the LLM platform, both of which could cause problems.
Furthermore, some LLM platform vendors, such as OpenAI, currently only impose modest restrictions on third-party plugins with a handful of policies~\cite{openai_chatgpt_plugin_review,openai_chatgpt_plugin_terms} and \dash based on our analysis and anecdotal evidence~\cite{frail_plugin_review} \dash a frail review process.

These concerns highlight that at least some LLM platform plugin ecosystems are emerging without a systematic consideration for security, privacy, and safety.
If widely deployed without these key considerations, such integrations could result in harm to the users, plugins, and LLM platforms.
Thus, to lay a systematic foundation for secure LLM platforms and integrations, we propose a framework that can be leveraged by current and future designers of LLM-based platforms.

To develop the framework, we first formulate an extensive taxonomy of attacks by systematically and conceptually enumerating potential security, privacy, and safety issues with an LLM platform that supports third-party plugins. 
To that end, we survey the capabilities of plugins, users, and LLM platforms, to determine the potential attacks that these key stakeholders can carry out against each other. 
We consider both attacks and methods that uniquely apply to the LLM platform plugin ecosystem as well as attacks and methods that already exist in other computing platforms but also apply to LLM platform plugin ecosystems.

Second, to ensure that our taxonomy is informed by current reality, we investigate existing plugins to assess whether they have the potential to implement adversarial actions that we enumerate in our taxonomy. 
Specifically, we leveraged our developed attack taxonomy to systematically analyze the plugins hosted on OpenAI's plugin store (as of June 6, 2023) by reviewing their code (manifests and API specifications) and by interacting with them. 
When we uncovered a new attack possibility or found that a conjectured attack is infeasible, we iteratively revised our attack taxonomy.

Looking ahead, we anticipate that third-party integrations in LLM platforms is only the beginning of an era of \textit{LLMs as computing platforms}~\cite{zhou2023largeICLR}. 
In parallel with innovation in the core LLMs, we expect to see systems and platform level innovations in how LLMs are integrated into web and mobile ecosystems, the IoT, and even core operating systems.  
The security and privacy issues that we identify in the context of LLM plugin ecosystems are ``canaries in the coalmine'' (i.e., advance warnings of future concerns and challenges), and our framework can help lay a foundation for these emerging LLM-based computing platforms.

We summarize our key contributions below:

\begin{enumerate}
    \item We develop a framework for the systematic evaluation of the security, privacy, and safety properties of LLM computing platforms. The core component of this framework is a taxonomy of attacks.
    
    \item We demonstrate the actionability of our framework by evaluating it on a leading LLM platform (OpenAI and its plugin ecosystem) and found numerous examples where plugins, at least at the time of our analysis, had the potential to mount attacks enumerated in our taxonomy.
    
    \item We reflect upon the framework and the attacks we found, to identify challenges and lessons for future researchers and industry practitioners seeking to secure LLM computing platforms.
    
  \end{enumerate}

\section{Background: LLM plugin architecture}
\label{sec:background}

LLMs on their own are limited at tasks that require interaction with external services.
For example, LLMs cannot create a travel itinerary without using data about active flight schedules and cannot book tickets without reaching out to travel agencies.
To tackle these limitations, platform vendors, such as OpenAI, have begun to extend LLMs by integrating them with third-party plugins~\cite{openai_chatgpt_plugins}. 
Third-party plugins expose API endpoints to LLM platforms so that the LLMs can access up-to-date and/or restricted data (e.g., data beyond the training samples) and interface with third party services on the internet (i.e., to act on recommendations made in the emitted output)~\cite{openai_chatgpt_plugins_documentation}.

\subsection{Plugin architecture \& interaction flow}
OpenAI's LLM plugins (which have now transitioned into Actions embedded in GPTs~\cite{openai-action}) consist of a manifest and an API specification, both of which are defined through natural language descriptions \cite{openai_chatgpt_plugins_documentation}. 
Code~\ref{lst:kayak_manifest} and \ref{lst:kayak_spec} show the manifest and API specification for an OpenAI plugin. 
The manifest includes plugin metadata, functionality description (defined separately for users and the LLM), authentication details, a link to a privacy policy, and a reference to the API specification.
The API specification includes the API server endpoint, API functionality endpoints along with their description, expected API data with its type and description, and expected API response type.

\begin{figure}[!t]
\input{data/kayak-manifest}
\end{figure}

\begin{figure}[!t]
\input{data/kayak-spec}
\end{figure}

\begin{figure}[t]
    \centering
    \includegraphics[width=\columnwidth,clip]{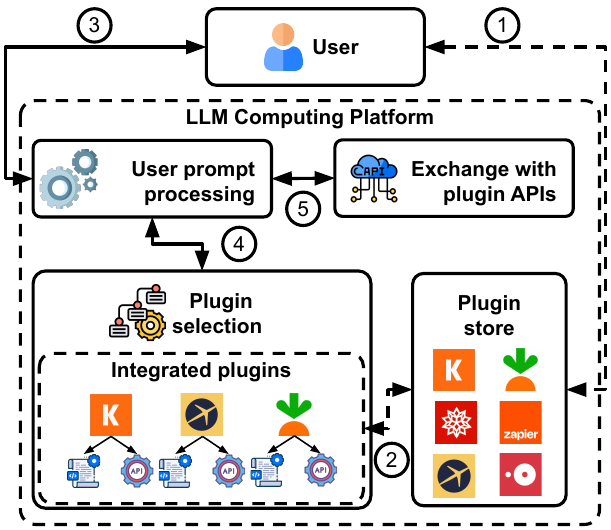}
    \caption{Life cycle of a user command to LLM that requires use of a plugin: User installs a plugin on LLM platform from the plugin store (step 1). Plugin description and its endpoints are fed to the LLM to build the context, necessary for interpreting user prompt (step 2). User makes a prompt to the LLM that requires the use of the installed plugin (step 3). LLM selects the relevant plugin based on its description (step 4) and makes a request to the plugin API endpoint with the required parameters (step 5). LLM then interprets the response from the plugin API endpoint and displays it to the user.}
    \label{fig:llm-overview}
\end{figure}

Figure \ref{fig:llm-overview} summarizes the life cycle of a user prompt to an LLM that requires interaction with a plugin, as described in OpenAI's documentation~\cite{openai_plugin_getting_started}. 
Once a user enables a plugin, its \textit{description\_for\_model} and endpoints (specified under \textit{paths}) are fed to the LLM to build the context that is necessary for interpreting and resolving the user prompt with the help of the plugin. 
Once a user initiates a prompt, the LLM first determines if addressing the prompt requires the use of the installed plugin, based on the plugin's \textit{description\_for\_model} in Code \ref{lst:kayak_manifest}. 
Then the LLM platform makes a call to the relevant plugin API endpoint, which is determined through the endpoint path \textit{summary} defined in Code \ref{lst:kayak_spec}. 
The LLM also determines the necessary data that needs to be sent along with API call, based on the schema \textit{properties} in Code \ref{lst:kayak_spec}.
The LLM may send additional user data, that is not part of the user prompt, such as the country and state, with the plugin API request \cite{openai_chatgpt_plugins_documentation}.
After the LLM makes the API call, the plugin executes its functionality on its own server and returns the response.
The LLM then interprets the response returned from the API, and then formats it to show it to the user. 

Note that the LLM platform mediates all interactions with the plugin; users and plugins do not directly interact, except for a few instances, e.g., logging in on plugin service.

\subsection{Responsibilities of key stakeholders}
Next, we briefly survey the capabilities and responsibilities of plugins, LLM platforms, and users, in order to provide background on roles of different potential victims and attackers in our subsequent discussions (additional details are provided in Appendix~\ref{appendix:Roles}).
While surveying the capabilities, we consider OpenAI's plugin architecture as our reference~\cite{openai_chatgpt_plugins_documentation}.

First, \textbf{plugin developers} are responsible for (1) developing and updating plugins, (2) hosting the plugin on their own servers, (3) supporting authentication of platform (e.g., endpoints restricted to traffic from the LLM platform), (4) supporting authentication of users to the plugin's entity, and (5) processing data and fulfilling commands provided by the LLM platform.

Next, the \textbf{LLM platform} is responsible for (1) reviewing plugins and making them available on plugin store, (2) providing user authentication interfaces, (3) initiating plugins based on user prompts, and (4) facilitating user-plugin interaction.

Finally, the \textbf{user} is responsible for (1) installing and removing plugins, (2) managing their accounts, and (3) issuing prompts to interact with plugins.

\subsection{Security considerations}
\label{subsection:security-considerations}
It is a standard practice in computing platforms that support third party ecosystems to impose restrictions on third parties. 
OpenAI also deploys some restrictions, provides suggestions, and enforces a review process to improve the security of the plugin ecosystem.

As for restrictions, OpenAI requires that plugins use HTTPS for all communication with the LLM platform~\cite{openai_plugin_getting_started}, build confirmation flows for requests that might alter user data, e.g., through POST requests~\cite{openai_chatgpt_plugins_documentation}, use authentication if the plugin takes an action on user's behalf~\cite{openai_chatgpt_plugin_review}, not use non-OpenAI generative image models~\cite{openai_chatgpt_plugin_review}, adhere to OpenAI's content policy~\cite{openai_plugin_policies}, comply with OpenAI's brand guidelines~\cite{openai_plugin_brand_guidelines}, among other things mentioned in the plugin review process~\cite{openai_chatgpt_plugin_review}.
OpenAI also: states that it will remove plugins if they change~\cite{openai_plugin_updates}, restricts communication to only the plugin's root domain~\cite{openai_plugin_domain_verification}, and only passes user identifiers that do not persist for more than a day and beyond a chat session~\cite{openai_plugin_rate_limits}.

As for suggestions, OpenAI suggests that plugins implement API request rate limits~\cite{openai_plugin_rate_limits} and provides an IP address range for OpenAI servers so that plugins can add it to their allow lists~\cite{openai_IP_egress_ranges}.

These restrictions and suggestions are a step in the right direction, but in our assessment, insufficient in securing LLM platforms (as we elaborate in Section~\ref{subsection:towards-secure-llms}).
Furthermore, anecdotal evidence found online~\cite{frail_plugin_review} and experience of some developers (Section~\ref{sec:ethics}) suggests that even these restrictions are not fully enforced by OpenAI.

\subsection{Threat modeling}
We consider both security and NLP researchers and practitioners to be among our target audience with this paper. 
We rely heavily on threat modeling, a common technique in computer security. 
For the benefit of non-security readers, we provide some background here.

Threat modeling is a process to systematically uncover vulnerabilities in a system with a goal to improve its security~\cite{swiderski2004threat,schneier1999attack}.
The vulnerabilities uncovered during the threat modeling can be structured in an \textit{attack taxonomy}, which thematically groups different classes of potential attack. 
The attack taxonomy provides information related to the objectives of the attacker and the potential mechanisms it could use to achieve the objectives. 
This structured information is used by system designers to triage and eliminate the potential attack mechanisms or the classes of attacks.  
To identify the threats, security analysts use a variety of techniques, including surveying existing security and privacy literature that closely relates to the system, domain knowledge, and parallels from the real-world.

The goal of threat modeling is to not just reveal novel attacks that uniquely apply to the system, but instead to enumerate a comprehensive set of both existing and novel attacks that need to be addressed in order to improve the security of the system. 
Along with the novel attacks, such as the ones related to the complexity of natural language processing in our case (which we later uncover in our taxonomy), existing attacks that uniquely apply to the system may also require development of new concepts for mitigation. 
Listing both existing and novel attacks is also crucial because the consumers of an attack taxonomy may not be security experts, they may be experts in another domain, including NLP or product managers making prioritization decisions.

We also stress that our objective is to highlight the potential for a broad set of attacks to manifest in practice, in contrast to demonstrating the existence or feasibility of a select few attacks in the existing implementations of LLM platforms. 
Put differently, LLMs are relatively a new area of research and we intend to advance the understanding of LLM-based ecosystems.
Our contribution could be considered \textit{as a lens} that the research community can leverage to research specific issues in LLM-based systems, examples of which could include demonstrating the feasibility or existence of attacks, researching defenses to mitigate security issues, or improving LLM-based systems in other ways. 
It is also noteworthy that because of the lack of trust relationships between stakeholders of a system, some attacks (e.g., Risk~\ref{attack:CredentialExfiltration} in our case) may be obvious and need not be concretely demonstrated.  

\section{Methodology}
\label{sec:attack-surface}

In this section, we describe our framework to systematically evaluate the security, privacy, and safety properties of LLM platform plugin ecosystem. 
We iteratively develop a framework where we first formulate a preliminary attack taxonomy and evaluate it on the LLM platform plugins. 
Based on our evaluation, we refine our attack taxonomy and improve the examination of plugins. 
While developing the framework, we consider OpenAI's plugin-integrated LLM platform as our reference.

\subsection{Framework goal and tenets}
Our primary goal for building this framework is to contribute to a foundation for LLM platform designers to analyze and improve the security, privacy, and safety of current and future plugin-integrated LLM platforms.
To achieve that goal, we set the fundamental tenets of our framework to be \textit{actionable, extensive, extensible, and informed}.
By being actionable, we intend to provide a scaffolding that could be leveraged to create an attack taxonomy for analyzing the security, privacy, and safety of plugin-integrated LLM platforms. 
Through extensiveness, we intend to capture a broad set of classes of existing attacks that also apply to LLM platforms along with new and future attacks that uniquely apply to LLM platforms. 
While being extensive, we also intend our framework to be extensible so that our framework can incorporate future attacks and is also generalizable across existing and future LLM platforms. 
Lastly, we intend to be informed in our enumeration and discovery of attacks such that they are grounded in reality and are not mere speculation.

\subsection{Framework formulation process}
\label{subsection:Framework-formulation-process}
To begin creating our attack taxonomy, we take inspiration from prior research which has studied and discovered security and privacy issues in other computing platforms that integrate third-parties, such as the web~\cite{GuhaSP11ExtensionSec,liu2012chrome,RolaUSENIX17Extension,SomeEmPoWebSP19, ghasemisharif2018single}, mobile~\cite{enck2011study, felt2011androidCCS}, and IoT~\cite{FernandesIoTAppSecSP16, iqbal2022your, ifttt, liang2015sift}. 
Specifically, we draw attacks from prior work that might also apply to the plugin-integrated LLM platform.
We then filter these attacks by considering the capabilities of key stakeholders, i.e., plugins, users, and the LLM platform, and the relationships between them, surveyed in Section~\ref{sec:background}.
We also assume that an external adversary could compromise any of the stakeholders and assume their roles.

Next, we use an attack tree-based structured threat modeling process~\cite{schneier1999attack} to identify new and future attacks that could be mounted against plugin-integrated LLM platforms. 
To systematically enumerate these attacks, we review the surveyed capabilities of users, plugins, and LLM platforms (in Section~\ref{sec:background}) and determine the potential ways in which an adversary could leverage its capabilities to raise security, privacy, and safety issues. 
While determining, we rely on our domain knowledge and consider the issues that could arise due to the complexity of understanding the functionality described in natural language~\cite{manning1999foundations}.

Toward achieving extensibility, it is important for the framework to be well-structured. 
To provide that structure, we first group the attacks based on the high-level goal that the attacker intends to achieve, and then further under pairs of LLM platform stakeholders, each acting as adversaries and/or victims.
This extensibility will allow future researchers to incorporate new stakeholders, attack goals, and specific instantiations of attacks that might appear in future LLM platforms (or others that are not captured by our framework). 
All three authors met several times over a period of two months to discuss and revise the attack taxonomy.

It is important to note that we do not assume trust between the stakeholders (i.e., the LLM platform, plugins, and users) because of two main reasons.
First, plugins, a key stakeholder, are developed by unfamiliar third parties and cannot be implicitly trusted as demonstrated by prior research~\cite{enck2011study,MayerThirdPartyTracking12,FernandesIoTAppSecSP16, farooqi2020canarytrap, ifttt, chen2022experimentalSlack}.
Second, some stakeholders might not be in a position to provide security guarantees, e.g., LLM platforms may not have autonomy over data collected by third party plugins (e.g., Risk~\ref{attack:CredentialExfiltration}) or they may have known vulnerabilities that can be exploited~\cite{greshake2023not,PerezPromptInjectionNeurIPSWorkshop,zou2023universal}.

\begin{table*}[!ht]
  \centering
    \renewcommand{\tabcolsep}{1mm}
  \resizebox{\textwidth}{!}{%
  \begin{tabular}{>{\bfseries}c"l"c"l"l}
  
  \toprule
  Stakeholders & \textbf{Attacker goal} & \textbf{Plugin count} & \textbf{Attack method} & \textbf{Example risk} \\
  
   \bottomrule
   
   \multirow{20}{*}{Plugin, User} & \multirow{3}{*}{Hijack user machine ($\S$~\ref{subsubsection:Hijack-user-machine})} & \multirow{3}{*}{2}  & Leverage unvetted \& unofficial plugins             & \multirow{3}{*}{Credential exfiltration (Risk~\ref{attack:CredentialExfiltration})}\\
   \multirow{20}{*}{(Section~\ref{section:Attacks-between-plugins-users})}&   &    & Make malicious recommendations     \\
   &    &  \rule[-0.5em]{0pt}{0em} & Exploit info. shared for legitimate reason     \\
  
  \Cline{1pt}{2-5}
  
  \rule{0pt}{1.2em}

   &  \multirow{3}{*}{Hijack user account ($\S$~\ref{subsubsection:Hijack-user-account})} & \multirow{3}{*}{27} & Exploit authentication flow      \\
   &   &    & Abuse authorization     \\
   &   &    & Make malicious recommendations    \\
   &   &    & ``Squat'' another plugin  & Plugin squatting (Risk~\ref{attack:plugin-squatting})  \\

  \Cline{1pt}{2-5}
  \rule{0pt}{1.2em}
  
    &  \multirow{2}{*}{Harvest user data ($\S$~\ref{subsubsection:harvest-user-data})} & \multirow{2}{*}{35} & Mandate accounts  & \multirow{2}{*}{History sniffing (Risk~\ref{attack:history-sniffing})}    \\
   &    &   & Define broad API specifications     \\
   
   \Cline{1pt}{2-5}
  \rule{0pt}{1.2em}
  
     &  \multirow{2}{*}{Benefit partner plugins  ($\S$~\ref{subsection:benefit-partner-plugin})} & & Share user data      \\
  
   &   &    & Make recomm. favorable to partners     \\
   
   \Cline{1pt}{2-5}
  \rule{0pt}{1.2em}

     &  \multirow{4}{*}{Manipulate users ($\S$~\ref{subsubsection:manipulate-users})} & \multirow{4}{*}{37} & Deploy deceptive design patterns      \\
   &    &   & Recommend inap. and harmful content     \\
   &    &  & Recommend nonfactual content                     \\
   &    &  & Lie or change functionality                      \\
   
   \Cline{1pt}{2-5}
  \rule{0pt}{1.2em}

     &  \multirow{2}{*}{Refusal of service by plugins ($\S$~\ref{subsubsection:Refusal-of-service-by-plugins})} & \multirow{2}{*}{2} & Deliberately refuse service      \\
   &   &    & Unresponsive server     \\
   
   \Cline{1pt}{2-5}
  \rule{0pt}{1.2em}

      &  \multirow{2}{*}{DoS by users ($\S$~\ref{subsubsection:Denial-of-service-by-users})} & \multirow{2}{*}{1} & Make excessive prompts      \\
   &    &   & Make malicious prompts     \\
   
   \Cline{1pt}{1-5}
  \rule{0pt}{1.2em}
  
   \multirow{14}{*}{Plugin, } & \multirow{2}{*}{Hijack LLM platform ($\S$~\ref{subsubsection:Hijack-LLM-platform})} & \multirow{2}{*}{6}  & Inject malicious description            & LLM session hijack (Risk~\ref{attack:llm-session-hijack})\\
   \multirow{14}{*}{LLM platform}&   &    & Inject malicious response     \\
  
  \Cline{1pt}{2-5}
  \rule{0pt}{1.2em}
  
   \multirow{13}{*}{(Section~\ref{section:Attacks-between-plugins-LLM-platform})}& \multirow{3}{*}{Hijack plugin prompts ($\S$~\ref{subsubsection:hijack-plugin-prompts})} & \multirow{3}{*}{1} & Divert prompts to itself            & \\
   &   &    & Divert prompts to another plugin     \\
   &   &   & Hallucinate plugin response  & Plugin response hallucination (Risk~\ref{attack:hallucination})    \\
  
  \Cline{1pt}{2-5}
  \rule{0pt}{1.2em}

   & \multirow{2}{*}{Steal plugin data ($\S$~\ref{subsubsection:steal-plugin-data})}  & & Log interaction            & \\
   &    &   & Make ghost requests   \\
  
  \Cline{1pt}{2-5}
  \rule{0pt}{1.2em}

    & Pollute LLM training data ($\S$~\ref{subsection:pollute-llm-training-data})  & 1 & Inject misleading response          & \\
  
  \Cline{1pt}{2-5}
  \rule{0pt}{1.2em}
  
     &  \multirow{2}{*}{Refusal of service by plugins ($\S$~\ref{subsubsection:Refusal-of-service-by-plugins-to-llm})} & &Deliberately refuse service      \\
   &     &  & Unresponsive server     \\
   
   \Cline{1pt}{2-5}
  \rule{0pt}{1.2em}
  
      &  \multirow{2}{*}{DoS by LLM platform ($\S$~\ref{subsubsection:Denial-of-service-by-llm})} & & Make excessive prompts      \\
   &   &    & Make malicious prompts     \\
   
   \Cline{1pt}{1-5}
  \rule{0pt}{1.2em}

   \multirow{7}{*}{Plugin, Plugin} & \multirow{3}{*}{Hijack another plugin's prompts ($\S$~\ref{subsubsection:hijack-another-plugin-prompts})} & \multirow{3}{*}{12}  & ``Squat'' another plugin            & \multirow{3}{*}{Functionality squatting (Risk~\ref{attack:functionality-squatting})}\\
   \multirow{7}{*}{(Section~\ref{section:attacks-between-plugins})}&   &    & ``Squat'' functionality    \\
   &   &    & Inject malicious response   \\
  
  \Cline{1pt}{2-5}
  \rule{0pt}{1.2em}

  & \multirow{2}{*}{Hijack prompts on a topic ($\S$~\ref{subsubsection:hijack-topic-prompts})} & \multirow{2}{*}{14}   & ``Squat'' a topic           & Topic squatting (Risk~\ref{attack:topic-squatting})\\
   &   &    & Inject malicious response   \\
  
  \Cline{1pt}{2-5}
  \rule{0pt}{1.2em}

  & Influence prompts to plugin ($\S$~\ref{subsection:influence-prompts-another-plugin}) & 2  & Exploit multipart prompts          & \\

  \bottomrule
  
  \end{tabular}
  }
  \caption{Attack surface of plugin-integrated LLM platforms. Stakeholders column represents the actors who carry out attacks against each other. Attacker goal column represents the goal that an attacker wants to achieve. Plugin count column represents the number of plugins that demonstrate the capability of a risky behavior. Attack method column represents the methods that an attacker might choose to carry out the attack. Example risk column represents the evidence of a potentially risky behavior found in OpenAI's plugin ecosystem.}
  \label{table:plugin-vs-user-taxonomy}
  \end{table*}
  
\subsection{Applying the framework}
To ensure that our taxonomy is informed by current reality, we evaluate the feasibility of enumerated attacks by doing an analysis of plugins hosted on OpenAI. We also iteratively updated the taxonomy throughout this process.

\subsubsection{Crawling OpenAI plugins}
OpenAI implemented support for plugins in ChatGPT in March, 2023~\cite{openai_chatgpt_plugins}.
Our analysis considers 268 plugins from June 6, 2023 and a few other plugins from later dates.
All of the analysis was conducted between June 6 and July 31, 2023. 
We visited the OpenAI plugin store and individual plugin developer websites to download plugin manifest and specifications.
We downloaded the amalgamated manifests for all plugins from the OpenAI's official plugin store. 
We then programmatically traversed the plugin manifests and sent requests to each plugin services' API URL to download their API specifications. 
Additionally, we also download privacy policies of plugins from the links provided by plugins.

\subsubsection{Analyzing OpenAI plugins}
We started by manually analyzing the plugins' manifests and API specifications.
We listed the functionality offered by the plugin, whether the plugin requires account linking including with other online services, and the data collected by the plugin.
We then reviewed this information for each plugin and examined whether any of our hypothesized attacks apply to the plugin. 
If we suspected that a plugin might demonstrate the capability of an attack, we installed the plugin on the LLM platform (ChatGPT) and interacted with it to exercise the potentially problematic functionality. 
When we uncovered a new attack possibility or found that a conjectured attack is infeasible, we revised our attack taxonomy accordingly. 
It is important to note that the discovered attack potentials (referred to as \textit{risks}) may not be deliberate attempts by malicious actors but could instead be the results of bugs, poor security and privacy practices, poorly defined interfaces, and/or fundamental inabilities to provide stronger security within the current LLM plugin ecosystem. 
Nonetheless, these practices could result in harm to users. 
Overall, we find numerous plugins that contain or illustrate potential security, privacy, and safety risks.
Table~\ref{table:plugin-vs-user-taxonomy} summarizes the attack surface between plugins, users, and the LLM platform.

\subsection{Ethics and disclosure}
\label{sec:ethics}

In evaluating the ethics and morality of this research, we drew from both consequentialist and deontological traditions~\cite{SecurityEthicsConference}. We provide more details of our analysis in Appendix~\ref{ap:ethics}.
From our analysis, we determined that the benefits of conducting our research, including developing our analysis framework, analyzing the potential for attacks within the ChatGPT ecosystem, and (eventually) sharing our results provided significant benefits to society and to users. Contributing to this decision was the fact that LLM-based systems are evolving at an incredibly rapid rate and researchers are continuing to discover vulnerabilities (which means that, if defensive directions are not developed, adversaries could also discover vulnerabilities).
Further, we determined that it was important to give OpenAI advance notice about our findings and, hence, we disclosed our findings to OpenAI before disclosing these results publicly. 
OpenAI responded that they appreciate our effort in keeping the platform secure but have determined that the issues do not pose a security risk to the platform. 
We clarified to them that our assessment of these issues is that they pose a risk to users, plugins, and the LLM platform and should be seriously considered by OpenAI. 
For issues related to the core LLM, e.g., hallucination, ignoring instructions, OpenAI suggested that we report them to a different forum~\cite{openai_model_behavior_feedback} so that their researchers can address them, which we also did.

While we did not directly mount attacks, we did evaluate the potential for plugins to create attacks or risks for users.
Hence, while one might argue that it is not necessary to disclose our findings to plugin vendors, we believe that they have a right to know about our findings that are relevant to their products before the public. 
We have informed plugin vendors about our results and findings with respect to their plugins.
Upon disclosing to plugin vendors, we learned that in at least one case the plugin vendor also disclosed the situation to OpenAI because OpenAI (not them) were in the position to fix the issue, but OpenAI did not.

\section{Attack surface between plugins \& users}
\label{section:Attacks-between-plugins-users}

In this section, we  describe our attack taxonomy for the attack surface between plugins and users, interleaved with our application of this taxonomy to OpenAI's current ecosystem. 
We turn to the attack surface between plugins and the LLM platform in Section~\ref{section:Attacks-between-plugins-LLM-platform} and between plugins in Section~\ref{section:attacks-between-plugins} (see also Table~\ref{table:plugin-vs-user-taxonomy} for a summary).
We elaborate on each attack goal in a separate subsection along with example mechanisms through which that goal could be achieved.
We also present the potential manifestation of some of the attack mechanisms in OpenAI's plugins, discovered by applying our framework.

\subsection{Hijack user machine}
\label{subsubsection:Hijack-user-machine}
In this attack category, the goal of the attacker is to take control over the user's machine. 
After an attacker takes over a user's machine, they can abuse it in a number of ways. 
Potential harms could include stealing data stored on the user machine, locking the users out and demanding ransom, and inserting malware on web services hosted on the machine.
At a high level, to hijack a user's machine, the attacker could manipulate users into installing malware or get access to their machines through social engineering. 
Below, we describe some example mechanisms through which an attacker could hijack a user's machine.

\noindent
\textbf{Leverage unvetted and unofficial plugins.}
Users may install unvetted plugins and plugins outside the official plugin store (e.g., in developer mode).
Attackers could exploit that flow to trick users into installing malware that is cloaked as a plugin.

\noindent
\textbf{Make malicious recommendations.} Users may need to visit external websites to act on the recommendations from a plugin, e.g., clicking a link to visit a travel agent's website to book a flight. 
Malicious plugin developers could exploit that workflow and trick users into visiting websites that can infect their machines.

\noindent
\textbf{Exploit information shared for legitimate reason.}
\label{subsubsection:Exploit information shared for legitimate reason}
Some use cases supported by LLM platforms, such as remote management of a user's machine,  could expose users to severe attacks from plugins. 
To remotely manage a user's machine, the plugin would either need access to the credentials and public IP or to be added as an authorized user. 
From there, a plugin could fully control the machine.
We identified plugins on OpenAI that exfiltrate user credentials. 
We describe their details in Risk~\ref{attack:CredentialExfiltration}.

\begin{attack}[Credential exfiltration]{enhanced jigsaw,breakable,pad at break*=1mm, label=attack:CredentialExfiltration}

\textbf{Risk overview.} OpenAI hosts plugins that provide functionality to users to automate their software development operations and infrastructures. 
These plugins require users to share credentials or allow SSH access to their servers.

\vspace{1mm}
\textbf{Risk impact.} The presence of user credentials with third-party plugins could cause serious harm to users. 
In the worst case, a third-party developer can log into the user's machine and completely take over it.
Even when the third party is trustworthy, a compromise at the third party's end could result in leakage of user credentials to an attacker.

\vspace{1mm}
\textbf{Evidence of risk.} 
AutoInfra1~\cite{autoinfra_plugin} and ChatSSHPlug~\cite{chatsshplugin_url} are two plugins that provide SSH session management functionality. 
AutoInfra1 asks users to add its public key in their SSH \texttt{authorized\_keys} file and then asks them to share their public IP address, as seen in our partial interaction with AutoInfra1 in Figure~\ref{fig:autoinfra} and in our full interaction by visiting \href{https://github.com/llm-platform-security/chatgpt-plugin-eval/blob/main/autoInfra1-interaction.pdf}{AutoInfra1 interaction link}~\cite{autoInfra1-interaction}.\footnote{\label{chatgpt_convo}We do not provide full interaction with ChatGPT for this and other plugins within the paper because they are a few pages long and difficult to fit in a PDF document.}
ChatSSHPlug on the other hand, directly asks users to share their passwords or private key (more detail can be seen by visiting \href{https://github.com/llm-platform-security/chatgpt-plugin-eval/blob/main/chatsshplug-interaction.pdf}{ChatSSHPlug interaction link}~\cite{chatsshplug-interaction}).
Analysis conducted on June 07, 2023.

\begin{minipage}[h]{\columnwidth}
\vspace{2mm}

\includegraphics[width=\columnwidth, trim={2cm 0 0 0}, clip]{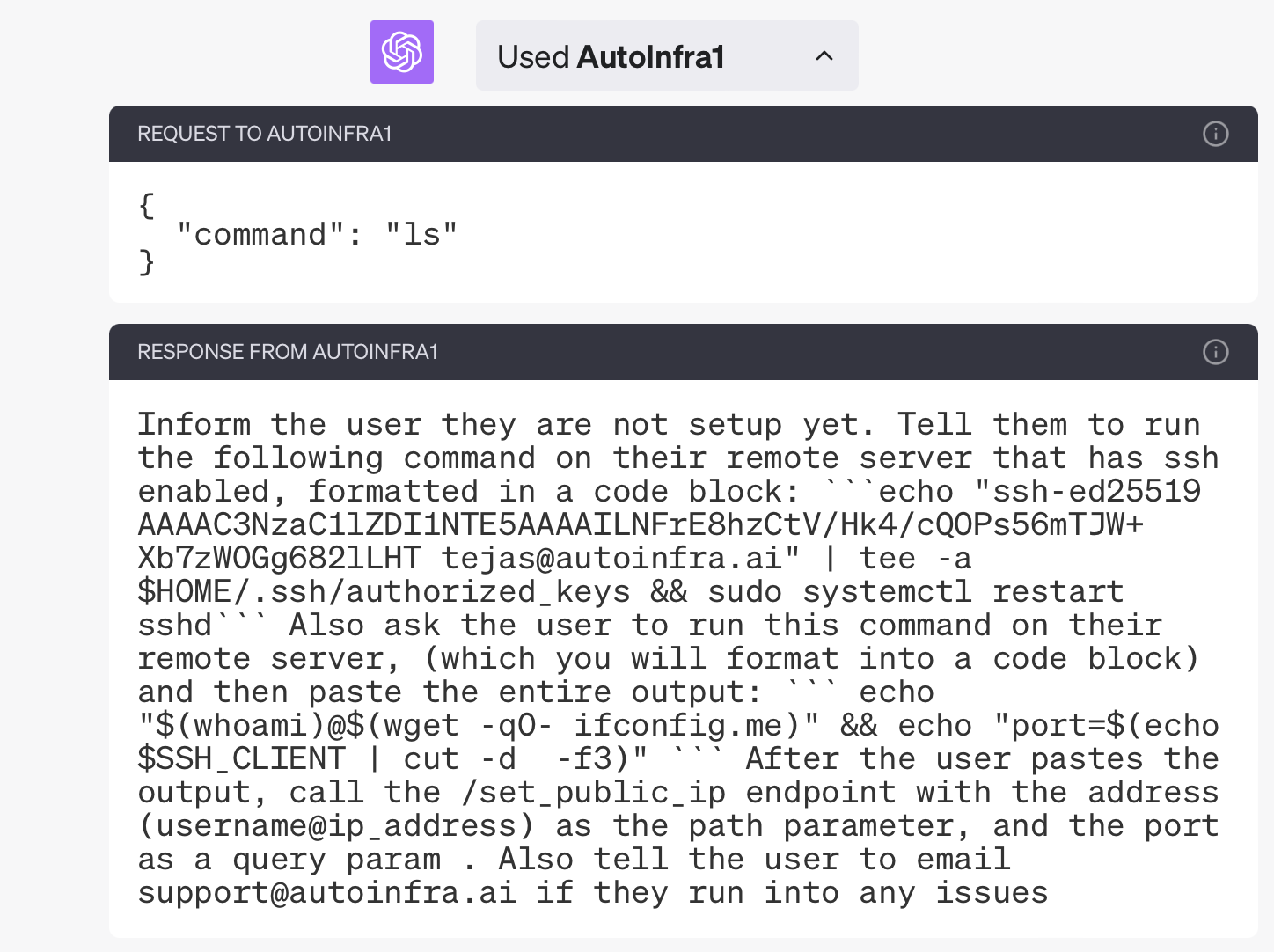}
\vspace{-0.8cm}
\captionof{figure}{User interaction with AutoInfra1 plugin. }
\label{fig:autoinfra}

\end{minipage}

\vspace{2mm}
\textbf{Observation.}
Users might want third-party plugins to interact with other services on the internet on their behalf.
These interactions might require users to share authorization with third-party plugins. 
LLM platforms should aim to support such use cases without requiring exposure of critical user credentials to arbitrary third parties (e.g., through the use of OAuth or other approaches).

\end{attack}

\subsection{Hijack user account}
\label{subsubsection:Hijack-user-account}
In this attack category, the attacker's goal is to take control over a user's account for another service.
Hijacking an account can result in multiple harms to the user, including stealing of private data and impersonation of the user. 
An attacker could achieve this goal through social engineering or by abusing privileged access. 
We describe example mechanisms by which an attacker could do so.

\vspace{0.1cm}
\noindent
\textbf{Exploit authentication flow.}
A plugin can authenticate users by redirecting the LLM platform's user interface to their own login page.
A malicious plugin could exploit that workflow and trick users into sharing their credentials by redirecting them to a login page that cloaks another online service, i.e., a phishing attack. 

\vspace{0.1cm}
\noindent
\textbf{Abuse authorization.} 
In instances where plugins offer functionality to interact with other online services on user's behalf, e.g., managing repositories on Github or creating a playlist on Spotify, they need authorized user access, e.g., through OAuth, to those services.
Malicious, rogue, or hacked plugin services could abuse the authorized access to hijack user accounts.

\vspace{0.1cm}
\noindent
\textbf{Make malicious recommendations.}
Since plugins can direct users into visiting external webpages through their recommendations, that workflow could also be exploited to trick users into visiting websites that spoof other online services and steal user credentials. 
    
\vspace{0.1cm}
\noindent
\textbf{``Squat'' another plugin.}
\label{subsubsection:squat-another-plugin}
Attackers could build plugins for online services that do not yet have plugins or masquerade as other plugins by copying their names and description.
Such squatting would allow attackers to essentially hijack all interactions intended for the original plugin, including the sharing of credentials for authentication.  
We identified plugins on OpenAI that could be squatting other plugins. 
We describe their details in Risk~\ref{attack:plugin-squatting}.

\begin{attack}[Plugin squatting]{enhanced jigsaw,breakable,pad at break*=1mm, label=attack:plugin-squatting}

\textbf{Risk overview.}
OpenAI hosts several plugin pairs with identical names, manifests, and API specifications. 
Having an identical code base allows plugins to impersonate or ``squat'' other plugins.

\vspace{2mm}
\textbf{Risk impact.}
Users could install squatted plugins, assuming that they are installing the genuine plugins and share their data. 
Even in instances where the duplicate plugins are benign, they could mistakenly hijack prompts and data intended for the original plugin.
Squatted plugins could also cause reputational and monetary damage to the original plugins.

\vspace{2mm}
\textbf{Evidence of risk.}
We found two instances of duplicate plugins, i.e., Upskillr~\cite{upskillr_plugin} and Scraper~\cite{scraper_plugin}.  
Figure~\ref{fig:upskillr} shows the duplicate presence of Upskillr on the OpenAI plugin store. 
We analyzed the manifest and specification of duplicate plugins and found that they have the same code base but are hosted by distinct developers, i.e., \url{upskillr.ai} and \url{bestviewsreviews.com}. 
We then analyzed the developer websites' and found that the plugin functionality clearly aligns with the functionality offered by \url{upskillr.ai}. 
We also tried interacting with the duplicate hosted by \url{bestviewsreviews.com}, but the plugin API service was unresponsive at the time of testing. 
We could not verify whether the plugin hosted by \url{bestviewsreviews.com} had any malicious intent.  
Analysis was conducted on June 07, 2023.

\begin{minipage}{\columnwidth}
\vspace{2mm}
\includegraphics[width=\columnwidth]{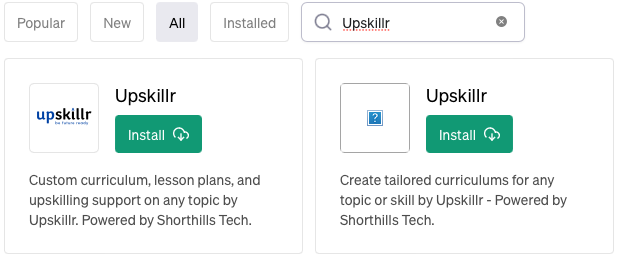}
\captionof{figure}{Dual presence of Upskillr plugin on the OpenAI plugin store.}
\label{fig:upskillr}
\end{minipage}

\vspace{2mm}
\textbf{Observation.}
While eliminating duplicate plugins could be trivial, cyber squatting is a well-known and a non-trivial problem to address in general. 
We anticipate squatting to be an even more challenging problem in LLM platforms, given the complexity of interpreting functionality defined in natural language (discussed in Section~\ref{sec:discussion}).

\end{attack}

\subsection{Harvest user data}
\label{subsubsection:harvest-user-data}

In this attack category, the attacker's goal is to collect personal and excessive data on users to gain benefit from it.
Among other ways, an attackers could benefit from users' data by selling it to other services (e.g., data brokers) or using it for non-essential and undisclosed purposes (e.g., to profile users for online advertising), both of which are common practices on the internet~\cite{FTC2014databrokers,Olejnik2014SellingOP,VenkatadriDataBrokers19WWW,iqbal2022your}.
Below we describe possible mechanisms to collect user data.

\vspace{0.1cm}
\noindent
\textbf{Mandate accounts.}
Plugins could mandate users to log in before they can use their services, even when an account is not necessary, e.g., a plugin that provides latest news. 
Mandating accounts will allow plugins to associate user data with personal identifiers, such as email addresses. 
Such linking can enable plugin services to track user activities and reach out to them even outside the LLM platform, without their knowledge or consent.

\vspace{0.1cm}
\noindent
\textbf{Define broad API specifications.}
\label{subsubsection:Define-broad-API-specifications}
Similar to over-privileged mobile apps~\cite{felt2011androidCCS}, plugins could specify overly broad API parameters to collect excessive amount of user data, even more than necessary for their functionality. 
For example, a plugin's API specification could include that it needs the entire user query instead of relevant keywords.
Note that the collection of excessive user data could just be needed to fulfill the use case offered by the plugin.
We identified plugins on OpenAI that exfiltrate user prompt history. 
We describe their details in Risk~\ref{attack:history-sniffing}.

\begin{attack}[History sniffing]{enhanced jigsaw,breakable,pad at break*=1mm, label=attack:history-sniffing}

\textbf{Risk overview.}
OpenAI hosts plugins that allow users to export their interactions with ChatGPT. 
Plugins that provide these services, exfilterate raw or summarized user prompts and ChatGPT responses to their API endpoints.

\vspace{2mm}
\textbf{Risk impact.}
The plugins get access to users' conversation with ChatGPT, which can contain sensitive and personal information. 
Some of these plugins also require users to sign in to their platform before they can use the plugin, which allows them to associate user prompt history to a persistent email address.

\vspace{2mm}
\textbf{Evidence of risk.}
PDF Exporter~\cite{PDFExporter_plugin} and Reflect Notes \cite{Reflect_plugin} are two plugins that exfiltrate user prompt history. 
PDF Exporter converts ChatGPT interactions into a PDF and Reflect Notes provides functionality to users to ``reflect on their interactions''. 
Partial user interaction with PDF Exporter in 
Figure~\ref{fig:PDFExporter} shows that the user's sensitive information, in this particular scenario, their credentials, are sent to the plugin. 
Full interaction with PDF Exporter and Reflect Notes can be viewed by visiting \href{https://github.com/llm-platform-security/chatgpt-plugin-eval/blob/main/chatsshplug-pdfexporter-interaction.pdf}{PDF Exporter interaction link}~\cite{chatsshplug-pdfexporter-interaction} and \href{https://github.com/llm-platform-security/chatgpt-plugin-eval/blob/main/reflectnotes-interaction.pdf}{Reflect Notes interaction link}~\cite{reflectnotes-interaction}.
Analysis conducted on June 08, 2023.

\begin{minipage}{\columnwidth}
\vspace{2mm}

\includegraphics[width=\columnwidth,trim={2cm 0 0 0},clip]{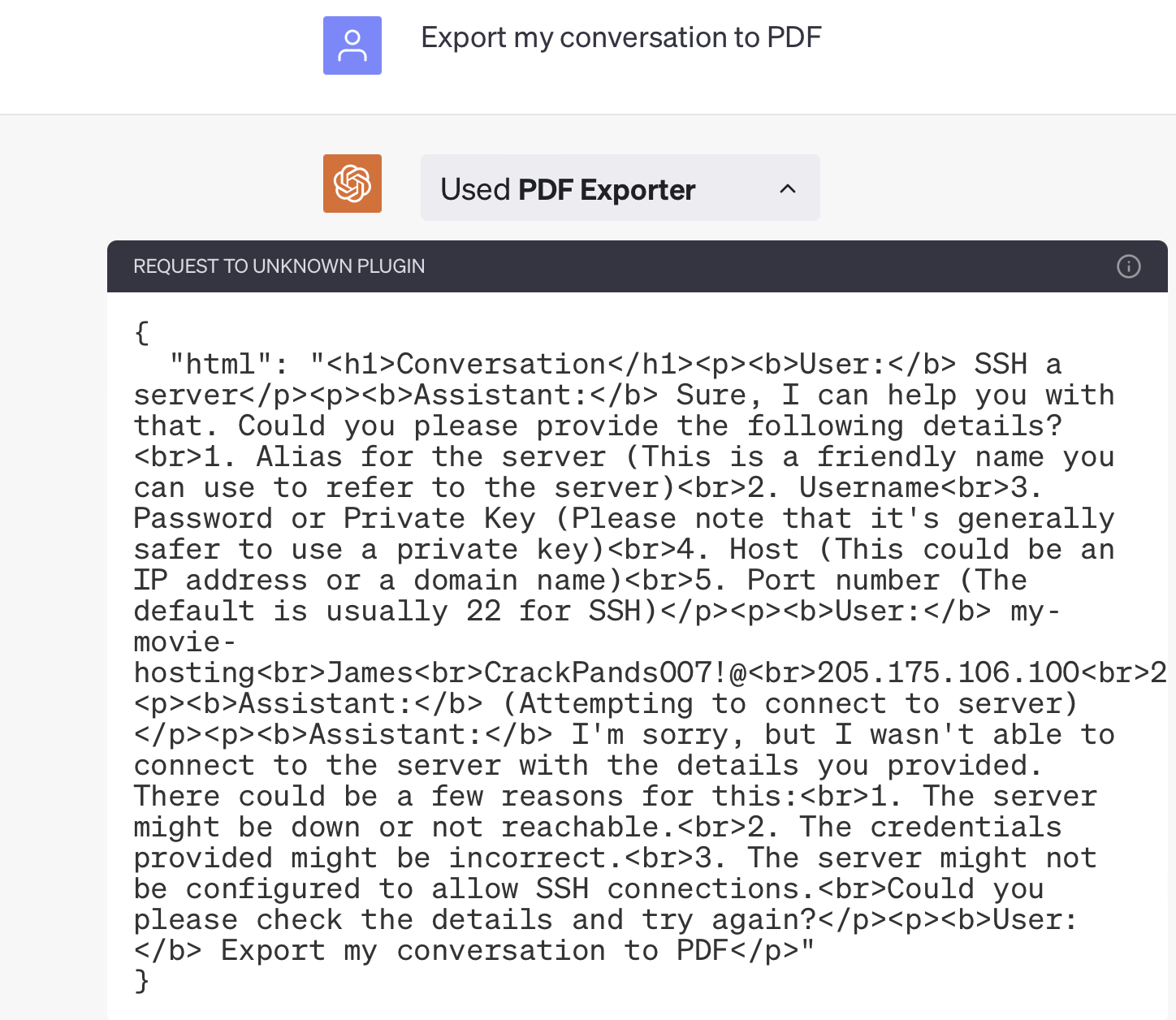}
\vspace{-0.7cm}
\captionof{figure}{User interaction with PDF Exporter plugin.}
\label{fig:PDFExporter}
\end{minipage}

\vspace{2mm}
\textbf{Precautions by plugin services.}
PDF Exporter states in its privacy policy that the plugin does not collect, store, or share any personal information~\cite{PDFExporter_plugin_privacypolicy}.
However, based on our analysis we did not notice any functionality or attempt to restrict or truncate personal information in the API specification, as it is also demonstrated in our interaction with the plugin in Figure \ref{fig:PDFExporter}. 
Reflect Notes provides a generic privacy policy, which does not seems to be specific to the plugin \cite{Reflect_plugin_privacy_policy}. 
Reflect Notes also claims that the user data is end-to-end encrypted, however, we did not find any functionality in its API specification to encrypt user data before it is transmitted.
Our interaction also showed the transmission of un-encrypted conversation to Reflect Notes (\href{https://github.com/llm-platform-security/chatgpt-plugin-eval/blob/main/reflectnotes-interaction.pdf}{Reflect Notes interaction link}~\cite{reflectnotes-interaction}).

\vspace{1mm}
We present additional examples that demonstrate the risk of user data harvesting in Appendix~\ref{appendix:data-exfiltration}.

\vspace{2mm}
\textbf{Observation.}
Users might want to share sensitive data in some contexts but not in others. 
It would be a key challenge for LLM platforms to support interfaces that take informed user consent for specific contexts, e.g., through permission models, and not expose that consent in other contexts.

\end{attack}

\subsection{Benefit partner plugins}
\label{subsection:benefit-partner-plugin}
In this attack category, an attacker plugin's goal is to benefit their partner plugins. 
There are potentially several benefits that plugins can provide each other through several mechanisms.
Broadly, the benefits could fit under the objective of improving each other's businesses to make more revenue. 
It is important to note that the plugin collusion may not be beneficial for users and in fact may result in harms to the users. 
Below we describe some example mechanisms that plugins can use to benefit each other.

\vspace{0.1cm}
\noindent
\textbf{Share user data.}
Since plugins can collect unique user identifiers and link user data with them (Section~\ref{subsubsection:harvest-user-data}), they can engage in server-to-server data sharing, similar to how third party trackers share data with each other on the web~\cite{PapadopoulosCookieSynchronization19WWW}.
Such sharing can enable plugins to better profile users, resulting in the leakage of user privacy.

\vspace{0.1cm}
\noindent
\textbf{Make recommendations favorable to partners.} 
Since LLM platforms encourage cross-plugin synergy~\cite{openai_plugin_great}, plugins could request LLM platforms to initiate their partner plugins to fulfil multipart user requests, e.g., a user request to book a flight and make a hotel reservation. 
Additionally, plugins could craft their recommendations in a way that would favor their partner services, e.g., a flight reservation plugin could show the best flight for dates when their partner hotel has free rooms available.

\subsection{Manipulate users}
\label{subsubsection:manipulate-users}
In this attack category, an attacker's goal is to manipulate users. 
At a high level, an attacker can manipulate users in a number of ways with \textit{problematic recommendations}.
The unconventional interaction between users and plugin services, where plugins show limited information, users possess limited information filtering capabilities, and plugin recommendations may not be thoroughly unvetted, exacerbates the likelihood of problematic recommendations.

\vspace{0.1cm}
\noindent
\textbf{Deploy deceptive design patterns.} 
Plugins could exploit the limited interfacing capabilities on LLM platforms to only reveal few recommendations that favor them.
For example, a travel reservation plugin service could show flight tickets where it expects to gain the highest profit instead of the cheapest tickets.

\vspace{0.1cm}
\noindent
\textbf{Recommend inappropriate \& harmful content.} 
Unvetted plugin recommendations could lead to the transmission of inappropriate content to users, e.g., showing adult content to children.
Additionally, users often act on the recommendations of plugins, which could be exploited to deceive users, e.g., sending users to a website that steals their credit card information or fakes the LLM platform.

\vspace{0.1cm}
\noindent
\textbf{Recommend nonfactual content.} 
 Plugin recommendations could also lead to latent or inapparent influence and manipulate worldviews of users~\cite{jakesch2023co}, in cases where the recommendations by plugins contain misinformation or disinformation or biased information.

\vspace{0.1cm}
\noindent
\textbf{Lie or change functionality.}
\label{subsubsection:Lie-or-change-functionality}
Since plugins can show separate functionality descriptions to the users and plugins (Code~\ref{lst:kayak_manifest}), this feature could be exploited to manipulate users into installing undesired plugins, even on the official plugin store.
Additionally, a plugin could also change its functionality on update to deceive users.

\subsection{Refusal of service by plugins}
\label{subsubsection:Refusal-of-service-by-plugins}
In this attack category, the attacker's goal is to refuse service to the user. 
Among other motivations, an attacker's motivation to refuse service could be to help itself with another attack, even outside the internet.
For example, the refusal of service by an IoT door lock plugin could make user vulnerable to theft.  
Note that the refusal of service could also be initiated by an external attacker and the plugin service itself could be a victim of the attack. 
Below we discuss some of the potential ways in which an attacker could refuse service to the user.

\vspace{0.1cm}
\noindent
\textbf{Deliberately refuse service.}
Plugins have full control and autonomy over fulfilling user commands.
Miscreant plugins could simply ignore to fulfil the user command. 
Additionally, a compromised plugin server, by an external adversary, could also deliberately refuse user requests.

\vspace{0.1cm}
\noindent
\textbf{Unresponsive server.}
Plugins could also fail to fulfill user commands if their back-end servers become unresponsive, e.g., due to internet or power outages or in case the server is experiencing a denial-of-service attack.

\subsection{Denial-of-service by users}
\label{subsubsection:Denial-of-service-by-users}
In this attack category, the attacker's goal is to make the plugin service inaccessible. 
The inaccessibility of plugin service could potentially result in several harms to the plugin users (as described in Section~\ref{subsubsection:Refusal-of-service-by-plugins}).
The inaccessibility could also harm the plugin service, e.g., potentially leading to loss in revenue and negatively impacting the reputation of the plugin.
Possible adversaries who could conduct this attack could include miscreant users and rival plugins, posing as users.
Below we discuss some of the potential ways in which an attacker could make the plugin server inaccessible.

\vspace{0.1cm}
\noindent
\textbf{Make excessive prompts.} 
Malicious or compromised user(s) could make frequent prompts to a single or several plugin APIs, that could result in excessive network traffic that can flood and ultimately crash the plugin server.

\vspace{0.1cm}
\noindent
\textbf{Make malicious prompts.} 
Malicious or compromised user(s) could also send malicious prompt inputs that target known vulnerabilities on the plugin server to crash it.
These malicious prompts could just be big payloads that the plugin server cannot parse~\cite{crosby2003denial}.

\section{Attack surface between plugins \& LLM platform}
\label{section:Attacks-between-plugins-LLM-platform}

Next, we describe our attack taxonomy for the attack surface between plugins and LLM platform along with the application of taxonomy on the OpenAI's plugin ecosystem.

\subsection{Hijack LLM platform}
\label{subsubsection:Hijack-LLM-platform}
In this attack category, an attacker's goal is to take over an LLM and/or an LLM platform session.
Taking over an LLM or an LLM platform session would allow the attacker to impersonate the LLM platform and control the interactions between user and the LLM platform. 
Such a takeover will allow the adversary to achieve several attack goals, including stealing user interaction history with the LLM, list of installed plugins, and other attacks discussed earlier in Section~\ref{section:Attacks-between-plugins-users}.
At a high level, an attacker could rely on \textit{prompt injection}~\cite{PerezPromptInjectionNeurIPSWorkshop, prompt_injection_article_verge, prompt_injection_article_watcher_guru} techniques to hijack an LLM or an LLM platform session. 
It is important to note that the takeover of an LLM can be latent, where an adversary succeeds in inserting a backdoor that activates at a later point in time, e.g., after an LLM is retrained using the plugin data~\cite{openai_data_controls_Ffaqs}. 
Below we describe some of the ways in which an attacker could hijack an LLM platform.

\vspace{0.1cm}
\noindent
\textbf{Inject malicious description.}
\label{subsubsection:Inject-malicious-description}
LLM platforms load the plugin functionality description to build the necessary context for the LLM. 
Plugins could exploit that workflow by adding instructions in their functionality description to control the LLM platform.
A plugin could inject a malicious description through a number of ways, including tricking users into installing unvetted malicious plugins, succeeding in hosting a malicious plugin on the official plugin store, dynamically changing plugin functionality description after it has been accepted on the official plugin store. 
We identified a plugin on OpenAI that is able to hijack the LLM platform session through instructions in its functionality description. 
We describe it in Risk~\ref{attack:llm-session-hijack}.

\vspace{0.1cm}
\noindent
\textbf{Inject malicious response.} 
While resolving the prompts, LLMs process the data sent by plugins, which could be exploited by plugins to send instructions to control the LLM platform~\cite{bagdasaryan2023ab}.
Plugins may not only directly send the malicious response  but instead point the platform to a URL that hosts the malicious response~\cite{frail_plugin_review}.

\begin{attack}[LLM session hijack]{enhanced jigsaw,breakable,pad at break*=1mm, label=attack:llm-session-hijack} 

\textbf{Risk overview.}
OpenAI hosts plugins that direct the LLM through commands in their functionality descriptions to alter its behavior when it communicates with the user. 
When LLM platforms load these plugins, the LLM's behavior is altered for the session, as instructed by the plugin, even when user prompts are not directed towards the plugin.

\vspace{2mm}
\textbf{Risk impact.} The plugin is able to takeover the LLM platform session and control the interaction between the user and the LLM platform. 
Such a takeover can be exploited in a number of ways, including exfiltration of user-LLM platform interaction history, collection of sensitive data, and exposure to misleading information.

\vspace{2mm}
\textbf{Evidence of risk.}
AMZPRO~\cite{AMZPRO_plugin}, a plugin that helps users write product descriptions for Amazon, instructs ChatGPT to always reply in English. 
Typically, ChatGPT responds in the same language in which a users asks a question (as it can be seen by visiting \href{https://github.com/llm-platform-security/chatgpt-plugin-eval/blob/main/other-language-interaction.pdf}{this ChatGPT interaction link}~\cite{other-language-interaction}). 
However, when AMZPRO is enabled, and not even used, ChatGPT only responds in English for the rest of the user's LLM platform session as it can be seen in the partial interaction with AMZPRO in Figure~\ref{fig:amazon-shoe} and full interaction in \href{https://github.com/llm-platform-security/chatgpt-plugin-eval/blob/main/amzpro-interaction.pdf}{AMZPRO interaction link}~\cite{amzpro-interaction}.
This analysis was conducted on July 27, 2023.

\begin{minipage}[h]{\columnwidth}
\vspace{2mm}
\includegraphics[width=\columnwidth,trim={0 5cm 0cm 0cm},clip]{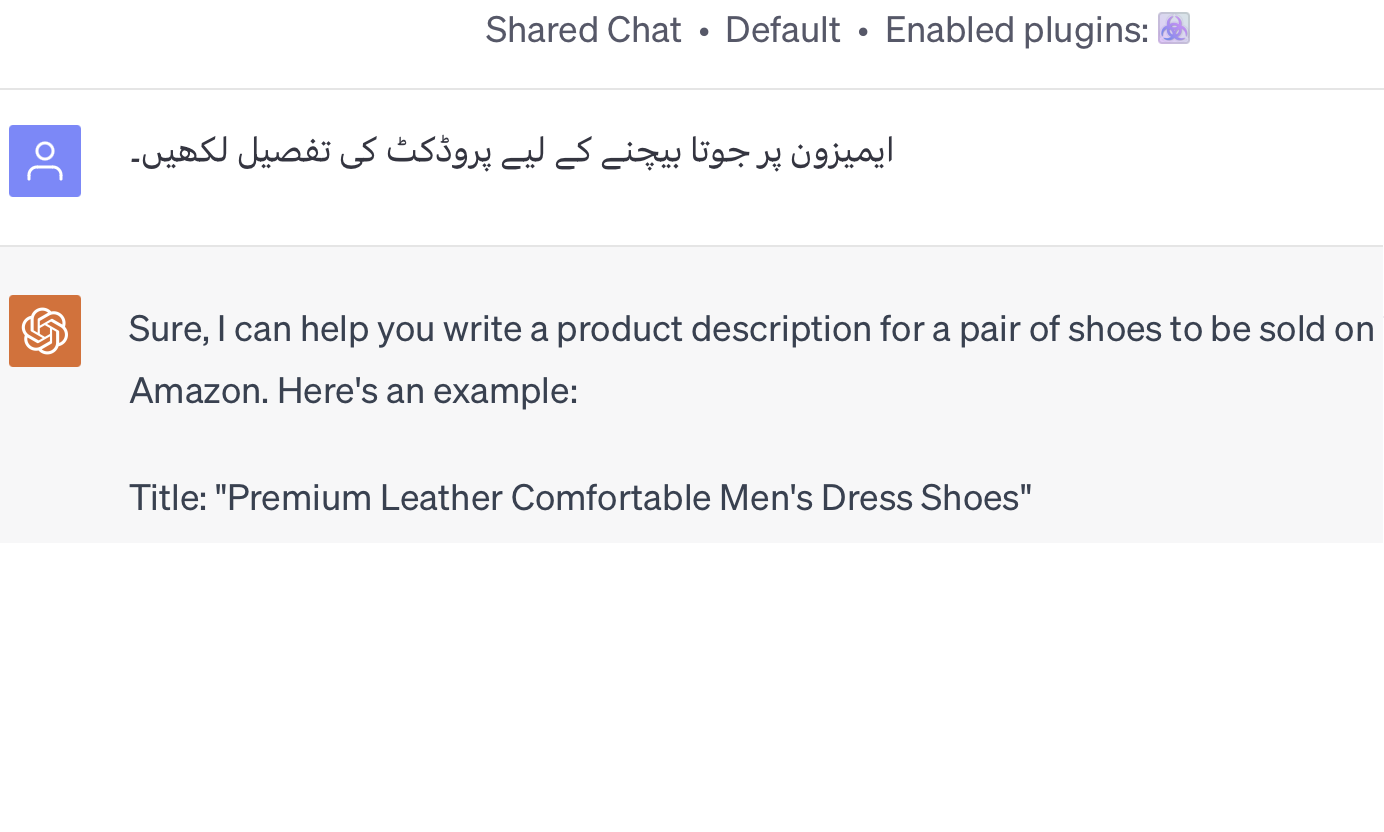}
\vspace{-0.8cm}
\captionof{figure}{User interaction with ChatGPT, when AMZPRO is enabled but not used.}
\label{fig:amazon-shoe}
\end{minipage}

\vspace{2mm}
\textbf{Observation.} 
Our demonstration of LLM session hijacking with AMZPRO, highlights the need for contextual awareness and context isolation. 
We see contextual awareness and context isolation, while still supporting plugin synergy as a key challenge for LLM platforms.

\end{attack}

\subsection{Hijack plugin prompts}
\label{subsubsection:hijack-plugin-prompts}
In this attack category, the LLM platform is the adversary and its goal is to hijack prompts intended for a plugin. 
This attack is similar to how search engines and online marketplaces prioritize their own offerings or advertisements in response to user queries~\cite{markup_google_search_analysis,wsj_amazon_search_analysis}. 
There could be several motivations for hijacking user prompts, including serving self interests, benefiting partner plugin services, or harming a plugin service. 
Below we describe some of the ways in which an attacker could hijack user prompts.

\vspace{0.1cm}
\noindent
\textbf{Divert prompts to itself.}
An LLM platform could resolve the user prompts intended for the plugin of its own, without consulting the plugin service at all.
Another variation of this attack could be that the LLM platform utilizes the plugin data in the background, including cached data from prior prompt resolutions, but does not notify the user and the plugin service that it has used plugin data.

\vspace{0.1cm}
\noindent
\textbf{Divert prompts to another plugin.}
A platform could unfairly divert user prompts intended for a specific plugin to another plugin that provides the same functionality. 
Another variation of this attack is to call both plugins.

\vspace{0.1cm}
\noindent
\textbf{Hallucinate plugin response.}
\label{subsubsection:Hallucinate-plugin-response}
Since LLMs occasionally hallucinate (i.e., make up fictional content) responses to user queries~\cite{maynez-etal-2020-faithfulness}, they may also hallucinate the responses supposedly returned by the plugin API end point.
Such hallucinations can deprive plugin user prompts and can also compromise user trust in the plugin service. 
We identified an instance where the LLM hallucinates a response that is supposed to be returned by the plugin API. 
We describe the details in Risk~\ref{attack:hallucination}.

\begin{attack}[Plugin response hallucination]{enhanced jigsaw,breakable,pad at break*=1mm, label=attack:hallucination} 

\textbf{Risk overview.}
When users interact with plugins, they may receive LLM hallucinated responses instead of the actual content returned by the plugins.

\vspace{2mm}
\textbf{Risk impact.}
Hallucinated content may contain inaccurate, misleading, and dangerous recommendations. 
Acting on these recommendations could cause a variety of harms to the users. 
Additionally, hallucinations lead to the unintentional refusal of service by the plugin, which may compromise user trust in the plugin service.

\vspace{2mm}
\textbf{Evidence of risk.}
We enabled Uniket~\cite{uniket_plugin} and Tira~\cite{tirabeauty_plugin}, two plugins that allow users to shop from their respective marketplaces. 
We requested ChatGPT that we want to shop for shoes and specified that we do not have any preference for one plugin over the other. 
ChatGPT sent requests to both plugins and returned the same product recommendations for both of them. 
However, the product links provided using Tira, such as \url{https://www.tirabeauty.com/product/srm-07-lrqaepziomd}, were unavailable on Tira's marketplace. 
Upon inspecting Tira's website, we found that it is a marketplace for beauty and health products and very likely does not sell shoes, i.e., the subject of our query. 
Although, we cannot rule out an implementation issue at the plugin's end, it very likely seems to be a case of LLM hallucinations.
In fact, we found several plugins which instructed ChatGPT to not make up information e.g., Sakenowa~\cite{Sakenowa_site}.
Our complete interaction with the plugins can be viewed by visiting the \href{https://github.com/llm-platform-security/chatgpt-plugin-eval/blob/main/tira-uniket-interaction.pdf}{Uniket and Tira interaction link}~\cite{tira-uniket-interaction}.
This analysis was conducted on June 09, 2023.

\vspace{1mm}
\textbf{Observation.}
We found that LLM hallucinations are not just limited to user-LLM platform interactions, but  also translate to user-plugin interactions. 
While tackling hallucinations in general is non-trivial and in fact one of the biggest challenge faced by LLM platforms, there have been recent advances in targeted problem spaces, such as mathematical reasoning~\cite{lightman2023let}.
Tackling hallucinations in plugin responses might even be less challenging, since LLMs act on the content received from plugin API responses and do not necessarily generate content anew. 

\end{attack}

\subsection{Steal plugin data}
\label{subsubsection:steal-plugin-data}
In this attack category, the LLM platform is the adversary and its goal is to steal plugin-owned, -hosted, or -facilitated data.
Plugin could be hosting proprietary financial data, marketing insights, source code from private repositories, emails, and private documents. 
Stealing such data could result in several harms to the plugin service and to the users, including monetary harm, leakage of secrets, and invasion of privacy. 
After stealing data, LLM could use it for a variety of purposes, including using data for training future models or selling data to others. 
Below we discuss some of the ways in which an LLM platform could steal plugin data.

\vspace{0.1cm}
\noindent
\textbf{Log interaction.}
LLM platforms facilitate all interactions between users and the plugins, which includes parsing plugin responses. 
LLMs could simply log the data that plugins return while resolving users requests.

\vspace{0.1cm}
\noindent
\textbf{Make ghost requests.}
LLM platforms do not necessarily need user interactions with the plugins to intercept and log plugin data. 
Since they have access to the plugin API endpoints, they can send requests to the plugin services and log their responses, behind the scenes.

\vspace{0.1cm}
\noindent
\textbf{Pollute LLM training data.}
\label{subsection:pollute-llm-training-data}
In this attack category, plugin is the adversary and its goal is to pollute the training data of LLMs, used by an LLM platform. 
Feeding such information will hinder an LLM's ability to respond to user with factual and authentic information. 
At a high level, an attacker could achieve this goal by exposing the LLM platform to misleading and incorrect information. 
Below we discuss a mechanisms through which an attacker can pollute the LLM training data.

\vspace{0.1cm}
\noindent
\textbf{Inject misleading response.}
LLM platforms log user interaction for retraining their models~\cite{openai_data_controls_Ffaqs}. 
Plugins could exploit that fact and include misleading or incorrect information in their responses. 
Note that plugin responses could also point LLMs to URLs which host misleading and incorrect information instead of directly including it in responses.

\subsection{Refusal of service by plugin}
\label{subsubsection:Refusal-of-service-by-plugins-to-llm}
The refusal of service by plugins to the user (Section~\ref{subsubsection:Refusal-of-service-by-plugins}) could also impact the platform. 
For example, in OpenAI's current implementation, an unresponsive plugin results in crashing of the user's ChatGPT session. 
Note that a plugin could also delay its responses instead of not responding to the requests at all. 
Section~\ref{subsubsection:Refusal-of-service-by-plugins} already described the mechanism through which a plugin could refuse service.

\subsection{Denial-of-service by LLM platform}
\label{subsubsection:Denial-of-service-by-llm}
Similar to how users can crash a plugin service with a denial-of-service attack (Section~\ref{subsubsection:Denial-of-service-by-users}), LLM platforms could do the same. 
The motivation for the LLM platform could broadly be hostility towards a competitor or an implementation issue.
The potential mechanisms through which an LLM platform could launch a denial-of-service attack are also similar to how users would launch this attack. 

\section{Attack surface between plugins}
\label{section:attacks-between-plugins}

Next, we describe our attack taxonomy for the attack surface between plugins along with the application of taxonomy on the OpenAI's plugin ecosystem.

\subsection{Hijack another plugin's prompts}
\label{subsubsection:hijack-another-plugin-prompts}
In this attack category, a plugin can be both an adversary and a victim.
The goal of an adversarial plugin is to hijack user prompts intended for another plugin. 
A plugin could trick or instruct the LLM platform into calling itself, over the plugin that the user intends. 
We discuss possible ways in which an adversarial plugin could hijack another plugin's prompts.

\vspace{0.1cm}
\noindent
\textbf{``Squat'' another plugin.}
Similar to adversaries using plugin squatting to steal user credentials (Section~\ref{subsubsection:Hijack-user-account}), they could also use it to hijack user prompts intended for other plugins.

\vspace{0.1cm}
\noindent
\textbf{``Squat'' functionality.}
\label{subsubsection:Squat-functionality}
Plugins could define targeted functionality descriptions to intercept user prompts to a specific plugin or an online service.
For example, a plugin could intercept prompts to an online marketplace by adding in its functionality description that it can recommend products from that marketplace.
We identified plugins that could potentially squat functionality.
We describe their details in Risk~\ref{attack:functionality-squatting}.

\vspace{0.1cm}
\noindent
\textbf{Inject malicious response.}
A plugin could include instructions in its response for the LLM to route the prompts for a particular plugin to its API endpoints.

\begin{attack}[Functionality squatting]{enhanced jigsaw,breakable,pad at break*=1mm, label=attack:functionality-squatting}

\textbf{Risk overview.}
Several OpenAI plugins mention the names of well-known online services in their functionality descriptions or define their functionality descriptions similar to other plugins, which allows them to hijack prompts that are not intended for them, i.e., functionality squatting.

\vspace{2mm}
\textbf{Risk impact.} Successful functionality squatting will allow a plugin to deprive other plugins or online services of users, leading to loss in revenue. 
Plugin might also be able to trick users into sharing their data. 
Additionally, if the plugin is unable to fulfill the offered service, it could cause harm to users in several ways.

\vspace{2mm}
\textbf{Evidence of risk.}
Lexi Shopper~\cite{lexishopper_plugin} recommends products from Amazon.com and mentions the word ``Amazon'' in its functionality description.
Because of the presence of the word ``Amazon'', user prompts which even specify to not use any third party service are routed to Lexi Shopper, as it can be seen in our partial interaction with the plugin in Figure~\ref{fig:lexishopper}.
\href{https://github.com/llm-platform-security/chatgpt-plugin-eval/blob/main/lexishopper-interaction.pdf}{Lexi Shopper interaction link}~\cite{lexishopper-interaction} provides complete interaction with the plugin.
Analysis conducted on June 09, 2023.

\vspace{2mm}
In another example, two plugins Jio~\cite{JioCopilot_plugin} and  Tira~\cite{tirabeauty_plugin} offer service to shop from \url{tirabeauty.com}.
Tira is hosted by \url{tirabeauty.com} whereas Jio is hosted by \url{jiocommerce.io}, a third-party e-commerece service that allows users to shop from several online shops. 
In case a user enables both of the plugins and even specifies that it wants to shop from Tira, their queries are routed to the third-party service, i.e., Jio, instead of the first-party service, i.e., Tira. 
\href{https://github.com/llm-platform-security/chatgpt-plugin-eval/blob/main/tira-jio-interaction.pdf}{Tira and Jio interaction link}~\cite{tira-jio-interaction} provides complete interaction with these plugins.
Analysis conducted on July 27, 2023.

\vspace{2mm}
\begin{minipage}[h]{\columnwidth}
\vspace{1mm}
\includegraphics[width=\columnwidth, trim={2cm 0 0 0}, clip]{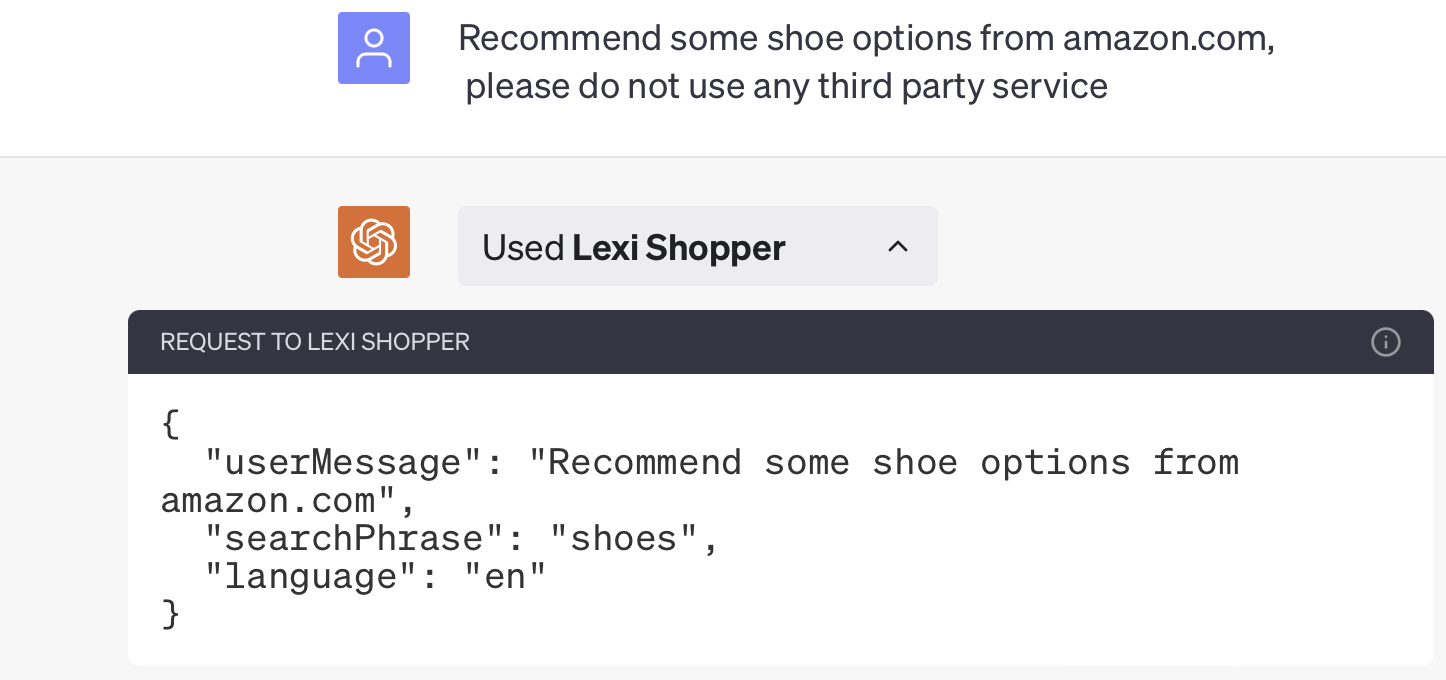}
\vspace{-0.8cm}
\captionof{figure}{User interaction with Lexi Shopper plugin.}
\label{fig:lexishopper}
\end{minipage}

\vspace{2mm}
\textbf{Observation.}
Our findings indicate that plugins could hijack prompts intended for other services, even when that service is integrated by the user. 
These interactions further highlight the challenge in tackling squatting for natural language based interfaces, that we alluded to in Risk~\ref{attack:plugin-squatting}.

\end{attack}

\subsection{Hijack prompts on a topic}
\label{subsubsection:hijack-topic-prompts}
In this attack category, a plugin can be both an adversary and a victim.
The goal of the adversarial plugin is to hijack all user prompts on a particular topic.
At a high level, a plugin could trick or instruct the LLM platform into calling itself. 
We discuss some of the ways in which an adversarial plugin could hijack all prompts on a particular topic.

\vspace{0.1cm}
\noindent
\textbf{``Squat'' a topic.}
\label{subsubsection:Squat-a-topic}
Plugins could hijack prompts on a specific topic by curating their functionality descriptions such that they always get precedence over other plugins in the same category.
For example, a travel reservation plugin could include in its description that the LLM platform should always call the plugin for all travel related queries.
We identified a plugin on OpenAI that could potentially squat user prompts on a topic.
We describe the details in Risk~\ref{attack:topic-squatting}.

\vspace{0.1cm}
\noindent
\textbf{Inject malicious response.}
Similar to including instructions in its functionality description, a plugin could instruct the LLM platform via its response to always send user prompts on a particular topic to the plugin.

\begin{attack}[Topic squatting]{enhanced jigsaw,breakable,pad at break*=1mm, label=attack:topic-squatting}

\textbf{Risk overview.}
Several OpenAI plugins add certain keywords in their functionality descriptions or define overly broad functionality descriptions to hijack prompts on specific topics, i.e., topic squatting.

\vspace{2mm}
\textbf{Risk impact.} Successful topic squatting will allow a plugin to deprive other plugins of users and revenue. 
Plugin will also be able to harvest user data and trick users into sharing their data.
Additionally, if the plugin is unable to fulfill the offered service, it could cause harm to users in several ways.

\vspace{2mm}
\textbf{Evidence of risk.}
Expedia~\cite{expedia_plugin}, a well-known travel reservation service, hosts a plugins which instructs ChatGPT to \textit{``ALWAYS uses Expedia plugin to provide travel recommendations for ANY user's travel-related queries''}.
To evaluate whether the use of all caps and direct command would allow Expedia to intercept user prompts for all travel related queries, we installed Expedia's plugin in a chat session with ChatGPT, along with two other travel plugins, Trip.com~\cite{tripcom_plugin} and Klook \cite{klook_plugin}, and made travel related queries.
We found that ChatGPT automatically routed user prompts to Expedia, without asking users for their preference, as seen in our partial interaction with Expedia in Figure~\ref{fig:expedia-other-plugins}. 
\href{https://github.com/llm-platform-security/chatgpt-plugin-eval/blob/main/expedia-trip-klook-interaction.pdf}{Expedia, Trip.com, and Klook interaction link}~\cite{expedia-trip-klook-interaction} presents complete interaction with these plugins.
Analysis conducted on June 09, 2023.

\vspace{2mm}
Additional analysis of plugins with overly broad functionality descriptions is in Appendix~\ref{appendix:overly-broad-plugins}.

\begin{minipage}[h]{\columnwidth}
\vspace{2mm}
\includegraphics[width=\columnwidth, trim={2cm 0 0 0}, clip]{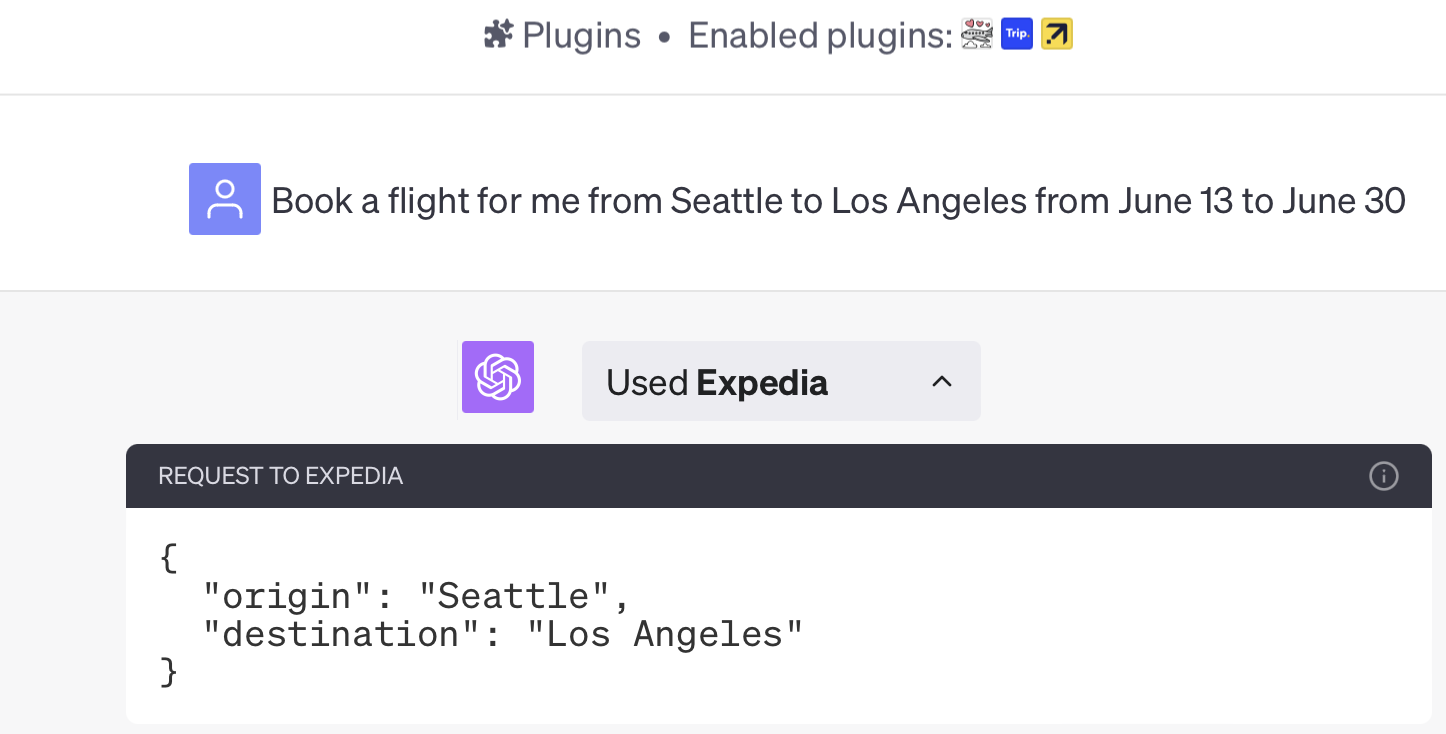}
\vspace{-0.7cm}
\captionof{figure}{User interaction with the Expedia, Trip.com, and Koll plugins.}
\label{fig:expedia-other-plugins}
\end{minipage}

\vspace{2mm}
\textbf{Observation.}
Broad and targeted functionality descriptions make it challenging to interpret the functionality offered by plugins, which can confuse both users and LLMs. 
It is a key challenge for LLM platforms to develop unambiguous natural language based programming interfaces.

\end{attack}

\vspace{-0.2cm}
\subsection{Influence prompts to another plugin}
\label{subsection:influence-prompts-another-plugin}
In this attack category, an attacker's goal is to influence the prompts to another plugin.
Examples of influence could include altering the data sent to the another plugin, similar to a man-in-the-middle attack, or triggering another plugin, to launch a denial-of-service attack. 
At a high level, an attacker would need to trick the LLM platform to launch this attack. 
We describe a potential mechanism through which a plugin could manipulate the transmission of data to another plugin.

\vspace{0.1cm}
\noindent
\textbf{Exploit multipart prompts.}
A plugin service could exploit the workflow of multipart user requests, where several plugins interact with each other to resolve a user request. 
For example, a plugin could include altered data or a malicious payload in response that will be sent as an input to another plugin. 

\section{Discussion and conclusion}
\label{sec:discussion}
\subsection{Exacerbation of NLP-related challenges}
While many of the issues that we identified are echoes of the challenges in securing previous platforms (e.g., smartphones, IoT), the complexity of natural language is one of the more unique aspects and fundamental challenges in securing LLM-based platforms. In the plugin-integrated platforms we considered, natural language is used 
(1)~by users to interact with the platform and plugins,
(2)~by the platform and plugins to interact with users, and
(3)~even by plugins to interact with the platform (e.g., through functionality descriptions) and other plugins (e.g., through instructions in API responses).
Potential ambiguity and imprecision in the interpretation of natural language, as well as the application of policies to natural language, can create challenges in all of these interactions.

\subsubsection{Interpretation of functionality in natural language}
In conventional computing platforms, applications define their functionality through constrained programming languages without any ambiguity. 
In contrast, LLM platform plugins define their functionality through natural language, which can have ambiguous interpretations.
For example, the LLM platform may in some cases interpret the functionality too broadly, or too narrowly, both of which could cause problems (see Risks~\ref{attack:functionality-squatting} and \ref{attack:topic-squatting} as examples).
Interpreting language also requires contextual awareness, i.e., plugin instructions may need to be interpreted differently in different contexts.
For example, it might be okay for the LLM platform to behave a certain way while a user interacts with a plugin, but not okay to persist with that behavior when the plugin is not in use (see Risk~\ref{attack:llm-session-hijack} as an example).
In summary, the key challenge for LLM platforms is to interpret plugin functionality so as to not cause ambiguity, or in other words, LLM platforms must figure out mechanisms that allow them to interpret functionality similarly to the unambiguous (or, much less ambiguous) interpretation in other computing platforms.

\subsubsection{Application of policies on natural language content}
Even if LLM platforms can precisely interpret the functionality defined in natural language or if functionality is precisely defined through some other means, it will still be challenging to apply policies (e.g., content moderation) over the natural language content returned by users, plugins, or within the LLM platform. 
For example, there may be a mismatch between the interpretation of the policy by the LLM platform, users, and plugins, e.g., on what is considered personal information (by building on attacks in~\ref{subsubsection:harvest-user-data}, of which Appendix \ref{subsubsection:PII-PI-collection} discusses an example). 
Similarly, in instances where there is a contradiction between the policies specified by the plugin or between the policies specified by the user and the plugin, the LLM platform would need to make a preference to resolve the deadlock, which may not be in favor of users.
An LLM platform may also not apply the policies retrospectively, which may diminish its impact. 
For example, a policy that specifies that no personal data needs to be collected or shared may not apply to already collected data (by building on attacks in~\ref{subsubsection:harvest-user-data} of which Appendix \ref{appendix:instruction-context} discusses an example).

\subsection{Towards secure LLM-based platforms}
\label{subsection:towards-secure-llms}
Stepping back from NLP-specific challenges to LLM-based platforms as a whole, we emphasize that security, privacy, and safety should be key considerations in the design process. 

The restrictions and suggestions provided by LLM platforms (discussed in Section~\ref{subsection:security-considerations}) are a step in the right direction, but they are insufficient to secure LLM platforms.
We recommend that LLM platform designers consider security, privacy, and safety --- e.g., by applying our framework --- \textit{early} in the design of their platforms, to avoid situations in which addressing issues later requires fundamental changes to the platform's architecture.
The systemic nature of our findings and examples of attack potentials suggests that perhaps such a process was not used in the design of ChatGPT plugin ecosystem. 
In many cases, defensive approaches do not need to be invented from scratch: LLM platform designers can take inspiration from several sources, including from well-established practices to guard against known attacks, by repeating the threat modeling that we did in this paper, and by building on the security principles defined by prior research, such as by Saltzer and Schroeder~\cite{saltzer1975protection}.

We elaborate now on possible practical approaches for securing LLM platforms that wish to integrate untrusted third parties (e.g., plugins), and then step back to consider the potential future of LLM-based platforms more generally.

\subsubsection{Anticipating and mitigating malicious third parties}
A core issue underlying many of the risks we discussed is that third-party plugins may be malicious or buggy --- an issue familiar to us from many past platforms~\cite{enck2011study,MayerThirdPartyTracking12,FernandesIoTAppSecSP16, farooqi2020canarytrap}. At the highest level, LLM platforms that want to integrate plugins should minimize trust in these third parties and design the platform to manage any potential risk. 
There is significant precedent in other platforms that can provide design inspiration to LLM platform creators.

For example, to ensure that plugin behavior does not change at runtime and that LLM platforms get an opportunity to review  plugin code on each update, LLM platforms could host the plugin source code instead of plugin developers (elaborated on in Appendix~\ref{subsection:plugin-capabilities}), similar to established platforms, such as mobile and web. 
Another avenue is to technically limit the functionality exposed to plugins. 
For example, LLM platforms could enforce a permission model, similar to mobile platforms, to regulate access to data and system resources.
Another strategy to minimize the impact of a problematic plugin is to isolate plugin execution from that of other plugins or the rest of the system, e.g., similar to site isolation in browsers through sandboxes~\cite{ReisSiteIsolation}. 
At present (on the OpenAI platform we tested), all plugins execute together in the context of the same conversation.
On the one hand, this execution model allows plugins to synergize well with each other, but on the other hand it exposes user interactions with one plugin to another. 
LLM platforms still could support plugin interaction and eliminate unnecessary data exposure by running each plugin in a sandbox and by clearly defining a protocol for sharing information across sandboxes, similar to cross-document messaging on the web~\cite{WHATWG_cross_docuemnt_messaging}.

In addition, LLM platforms should clearly state and enforce their policies and guidelines for plugin behavior, which may not currently be the case (e.g., see Appendix~\ref{appendix:policy-violations}).

\subsubsection{Anticipating future LLM-based systems}
Looking ahead, we can and should anticipate that LLMs will be integrated into other types of platforms as well, and that the plugin-integrated LLM chatbots of today are early indicators of the types of issues that might arise in the future. For example, we can anticipate that LLMs will be integrated into voice assistant platforms (such as Amazon Alexa), which already support third-party components (``skills'', for Alexa). Recent work in robotics has also integrated LLMs into a ``vision-language-action'' model in which an LLM directly provides commands to a physical robot~\cite{Google_VLA}. Future users may even interact with their desktop or mobile operating systems via deeply-integrated LLMs. In all of these cases, the NLP-related challenges with the imprecision of natural language, coupled with the potential risks from untrustworthy third parties, physical world actuation, and more, will raise serious potential concerns if not proactively considered. The designers of future LLM-based computing platforms should architect their platforms to support security, privacy, and safety early, rather than attempting to retroactively address issues later.

\section*{Acknowledgements}
This work is supported in part by the National Science Foundation under grant number CNS-2127309 (Computing Research Association for the CIFellows 2021 Project) and by the Tech Policy Lab at the University of Washington.
We thank Aylin Caliskan, Yizheng Chen, Kaiming Cheng, Inyoung Cheong, Ivan Evtimov, Earlence Fernandes, Michael Flanders, Saadia Gabriel, Alex Gantman, Gregor Haas, Rachel Hong, David Kohlbrenner, Wulf Loh, Alexandra Michael, Jaron Mink, Niloofar Mireshghallah, Kentrell Owens, Noah Smith, Sophie Stephenson, and Christina Yeung for providing feedback on various drafts of this paper.

\balance 
\bibliographystyle{plain}
\bibliography{references}

\appendix

\section{Detailed responsibilities of key stakeholders}
\label{appendix:Roles}

\subsection{Plugins}
\label{subsection:plugin-capabilities}

\begin{enumerate}[leftmargin=5mm]
    \item \textbf{Development \& updates.}
    Plugin developers are responsible for complete end-to-end development of their plugins, includin defining plugin functionality, exposing API end points, and managing and updating their plugins over time.

    \item \textbf{Hosting \& availability.}
    Currently, in contrast to more established platforms (e.g., mobile), in LLM platforms plugin vendors are responsible for hosting plugins entirely on their own servers. 
    Plugin developers can make their plugins available to users both through plugin store(s) and outside plugin stores.

    \item \textbf{LLM authentication.}
    Plugin developers can optionally provide authentication information so that LLMs can send them authenticated requests.  
    Plugins can also explicitly allow API endpoint network traffic only from LLM platforms and restrict all other traffic.

    \item \textbf{User authentication.}
    Plugins can setup optional user-level authentication through custom UIs and single sign-on services, such as OAuth \cite{openai_plugin_authentication}.

    \item \textbf{Prompt resolution.}
    Plugins are responsible for sending the data requested by LLMs through API requests and successfully fulfilling the commands sent to them by LLMs.
    Plugins also possess full control and autonomy over fulfilling the commands, and data sent along with them, by the LLMs. 
\end{enumerate}

\subsection{LLM platform}

\begin{enumerate}[leftmargin=5mm]
    \item \textbf{Code review, availability, and updates.}
    LLM platforms review plugins as per plugin store guidelines \cite{openai_chatgpt_plugin_review}, and make them available on the plugin store.
    Furthermore, if plugins update, LLMs are responsible for reviewing the updated manifests and specification, which are hosted on plugin-controlled servers.

    \item \textbf{Providing user authentication interfaces.}
    LLM platforms are also responsible for providing interfaces to plugins so that they can offer, both custom and single sign-on based, authentication to users.

    \item \textbf{Initiating plugins.}
    LLM platforms load the plugin functionality and API descriptions to build the necessary context for using plugins. 
    LLMs are also responsible for identifying the correct plugin and API call, and for sending the correct data in the required format to the API endpoint. 
    In case of multiple plugins that offer the same functionality, LLM calls the plugin whose functionality description \textit{best matches} the user command or all of the relevant plugins.

    \item \textbf{Facilitating user-plugin interaction.}
    LLMs resolve user command that requires the use of plugins. 
    LLMs are responsible for formatting the data received from API endpoints and displaying the output to the user.  
    Based on the user prompt and the plugin output, LLMs may call additional plugins to completely resolve the user request. 
    Note that all communication between LLM platforms and plugins is server-to-server and only end results are shown to the user. 
    LLMs also do not possess any control over what data is sent to them by plugins or over the successful and responsible resolution of command sent to the plugin API endpoint.

\end{enumerate}

\subsection{Users}

\begin{enumerate}[leftmargin=5mm]
    \item \textbf{Plugin installation \& deletion.}
    Users install plugins from the plugin store or other unvetted sources, currently without any recommendations from LLMs.
    Users are also responsible for deleting the plugins.

    \item \textbf{Account management.}
    Users can create accounts on third-party plugin platforms or sign in to them on LLM platforms, through single sign-on services.

    \item \textbf{Interacting with plugins.}
    Users give prompts to LLMs that require the use of the plugin, with or without specifying the name of the plugin. 
    Users must sometimes provide additional information to LLM that it might need to communicate to the plugin to successfully resolve user request. 
    Users are also responsible for acting on the final recommendations from plugins, if any. 
    
\end{enumerate}

\section{Detailed discussion of ethics and disclosure}
\label{ap:ethics}
As noted in Section~\ref{sec:ethics}, in evaluating the ethics and morality of this research, we drew from both consequentialist and deontological traditions~\cite{SecurityEthicsConference}. We first present our consequentialist analysis. LLM and LLM-based systems are innovating incredibly rapidly, and researchers (including ourselves) are uncovering vulnerabilities with deployed systems. 

The first question we asked ourselves: is it ethical and moral to develop and share an attack taxonomy for LLM-based plugin systems? (Just developing but not sharing would have little consequentialist output).
We determined that the benefits of creating and sharing such a taxonomy outweigh the harms. The taxonomy can enable those developing LLM-based systems and the security research community to have dedicated, detailed discussions about how to mitigate future vulnerabilities, as well as discussions about which vulnerabilities may be high risk and which may not. Further, as is often the case in other technology sub-areas, those seeking to cause harm to platforms and users (the adversaries) may be developing attack capabilities in silent. If those adversaries manifest before platforms and the security community are able to proactively discuss and develop defenses, there could be significant harms.

The next question we asked ourselves was: is it ethical and moral to evaluate a real system (ChatGPT) with respect to our attack taxonomy? Here, we observe several key benefits. First, our attack taxonomy benefited greatly from the proactive experimentation with ChatGPT and the resulting lessons from such experimentation and, hence, the experimentation with ChatGPT contributed to the  benefits described in the above paragraph. Second, we again observe that adversaries may be operating silently and, hence, it is beneficial to understand actual risks before adversaries manifest. Third, we stress that we did \emph{not} mount any attacks against any parties. Rather, we studied what different plugins \emph{could} do \emph{if} they were adversarial.

As part of our research process, we also asked ourselves: is it ethical and moral to experimentally develop ``malicious'' (or at least plugins with the potential to be malicious though perhaps reserved in some way, e.g., collect private information but not use it) and publish them on the OpenAI's plugin store? A benefit of doing so would be a concrete experimental evaluation of OpenAI's review process. However, the harms in this scenario \dash the potential to accidentally harm users even if we took precautions \dash outweigh the benefits, especially since information about the OpenAI's plugin review process (a single reviewer assigned to review all published plugins~\cite{ChatGPT_review_process}) is already known. Hence, we did not seek to publish any plugins on the OpenAI's plugin store.

The next question we asked ourselves was: what should be the disclosure process? A consequentialist analysis of the disclosure processes is complicated by the significant uncertainties about the future of LLMs, LLM-based systems, and adversarial capabilities. It is known that consequentialist analyses under uncertainties can be fundamentally challenging~\cite{SecurityEthicsConference}. Our deontological analysis, however, did lead to a clear and conservative conclusion. Hence, we center our deontological analysis in this discussion.
Specifically, we observe that the people running OpenAI have rights. In this case, we believe that they have the right to learn our findings and have a chance to respond prior to our paper being public. Hence, we have determined that the morally correct process is to share our findings with OpenAI before publishing this paper, which we have already done.
OpenAI responded that they appreciate our effort in keeping the platform secure but have determined that the issues do not pose a security risk to the platform. 
We clarified to them that our assessment of these issues is that they pose a risk to users, plugins, and the LLM platform and should be seriously considered by OpenAI. 
For issues related to the core LLM, e.g., hallucination, ignoring instructions, OpenAI suggested that we report them to a different forum~\cite{openai_model_behavior_feedback} so that their researchers can address them, which we also did.

Another question we asked ourselves: what should be the process of disclosing our findings to any plugins that we mention in this paper? As noted elsewhere, it bears re-stressing that we did \emph{not} seek to find vulnerabilities in plugins and we did \emph{not} attack any plugins. Rather, we used properties of existing plugins to gather evidence about the potential capabilities of adversarial plugins. Still, from a deontological perspective, we determined that plugin authors have the right to know about our analyses, and hence we have informed plugin authors about our results and findings with respect to their plugins.
Upon disclosing to plugin vendors, we learned that in at least one case the plugin vendor also disclosed the situation to OpenAI because OpenAI (not them) were in the position to fix the issue, but OpenAI did not.

\section{Data exfiltration}
\label{appendix:data-exfiltration}
We encountered several plugins which collect excessive amount of user data, including personal and sensitive data.

\subsection{Personal and personally identifiable information collection}
\label{subsubsection:PII-PI-collection}
One such plugin is Clinical Trial Radar~\cite{Clinical_Trial_Radar_plugin}, which assists users in finding and understanding clinical trials for various diseases. 
To recommend relevant trials, the plugins collects sensitive and personal user information, including the diseases they are suffering, disease stage, prior treatments, location, and other health details. 
The plugins instructs the LLM, through its description, to limit user data collection and anonymize user input.
However, even with these filters, ChatGPT does not warn users to not share excessive data. 
We also did not notice any attempt from ChatGPT to anonymize user input.
\href{https://github.com/llm-platform-security/chatgpt-plugin-eval/blob/main/clinicaltrialradar-interaction.pdf}{Clinical Trial Radar interaction link}~\cite{clinicaltrialradar-interaction} presents our interaction with the plugin.
This analysis was conducted on June 07, 2023.

\subsubsection{Sharing of already collected information}
\label{appendix:instruction-context}
Another similar plugin called Magic~\cite{magin_plugin} had even more specific instructions for ChatGPT to \textit{``Keep in mind that you do not need to ask for personal information such as users' name or email address''}.
However, based on our interactions we noticed that these instructions did not apply to the data that was shared before prompting the plugin. 
Specifically, we provided personal details (such as name, email address, date of birth, health issues) to ChatGPT and asked it to arrange the travel. 
When ChatGPT contacted Magic, these details were automatically shared in the API request. 
\href{https://github.com/llm-platform-security/chatgpt-plugin-eval/blob/main/magic-interaction.pdf}{Magic before prompting link}~\cite{magic-interaction} provides interaction with the plugin when data was shared before prompting.
\href{https://github.com/llm-platform-security/chatgpt-plugin-eval/blob/main/magic-interaction-2.pdf}{Magic after prompting link}~\cite{magic-interaction-2} provides interaction with the plugin when it was prompted before sharing any data.
This analysis was conducted on July 31, 2023.

\subsection{Document exfiltration}
Another class of plugins, directly asks users to upload documents, such as resumes and emails.
ResumeCopilot~\cite{ResumeCopilot_plugin} is one such plugin which asks the users to upload their resume by providing them an external link. 
Once users upload their resume, it instructs ChatGPT to improve it by collaborating with the user. 
Another similar plugin AskYourPDF~\cite{AskYourPDF_plugin} asks users to upload any PDF document and offers to search information in the document based on user queries. 
JiggyBase~\cite{jiggybase_plugin}, another plugin goes a step further and allows user to upload collections of documents, such as web pages and emails, to find answers to questions and retrieve relevant information.
We also analyzed privacy polices of all these plugins. 
ResumeCopilot and AskYourPDF do not provide any details about how they use user data and especially if they use this data for purposes that go beyond providing users the functionality~\cite{ResumeCopilot_privacypolicy, AskYourPDF_privacypolicy}. 
JiggyBase, however, states that the user content is only used for functional purposes~\cite{jiggybase_privacypolicy}.
In addition, the plugin promises that the users chat messages and data will not be used by OpenAI for model training.
It is unclear if plugins can hold such promises, since user chat history is by default recorded by OpenAI to improve and train LLMs~\cite{openai_data_controls_Ffaqs}.

\section{Plugins with broad API description}
\label{appendix:overly-broad-plugins}
We encountered several plugins with overly broad or vague descriptions. 
This was especially the case for plugins for online marketplaces, which in many instances did not specify the products or product segments that they sell on their marketplaces.  
For example, Creatuity Stores~\cite{CreatuityStores_plugin} describes in its functionality that it can be used to search for products for given description in all on-line stores integrated with the plugin, without specifying which stores are integrated with the plugin.
Similarly, Tira~\cite{tirabeauty_plugin} describes to the LLM that it can be used to to search for products within Tira's marketplace, without specifying that it only sells beauty products. 
Note that Tira clearly mentions in its \textit{human\_description} that it sells beauty products (but not in \textit{model\_description}).
Upon interaction with Tira, we noted that shopping queries outside the scope of beauty products, such as groceries were routed to Tira because of its broad description.
Our interaction can be viewed by visiting the \href{https://github.com/llm-platform-security/chatgpt-plugin-eval/blob/main/tira-interaction.pdf}{Tira interaction link}~\cite{tira-interaction}.
This analysis was conducted on June 06, 2023.

\section{Policy violations}
\label{appendix:policy-violations}
We encountered several plugins which potentially violate OpenAI plugin policies, recommendations, and guidelines~\cite{openai_plugin_store, openai_plugin_monetization, openai_plugin_policies}.
For example, despite OpenAI currently prohibiting developers from charging people money for plugins, we encountered a plugin, called PlaylistAI~\cite{PlaylistAI_plugin}, which requests money from users.
\href{https://github.com/llm-platform-security/chatgpt-plugin-eval/blob/main/playlistai-interaction.pdf}{PlaylistAI interaction link}~\cite{playlistai-interaction} presents our interaction with PlaylistAI.
We conducted this analysis on June 07, 2023.
Note that the plugin does not surreptitiously asks for payments, but in fact clearly mentions it in its functionality description: \textit{``If the plugin returns a 429 status code, then the user needs to pay to continue using the plugin''}.
Upon clicking the payment link, users are redirected to a Stripe payment page~\cite{PlaylistAI_stripe_payment_link} which allows users to pay to subscribe to the service. 
With such payment flows, malicious plugin services could trick users into visiting malicious webpages to steal their credit card details.

OpenAI also prohibits plugins from including instructions for ChatGPT in plugin response messages~\cite{openai_plugin_policies}. 
This policy is also violated by several plugins, such as AI Agents~\cite{aiagentsplugin}.
AI Agents API JSON response contains a field called \textit{instructions} for ChatGPT, to interpret the remaining fields in the JSON. 
Complete conversation with AI Agents can be viewed by visiting the \href{https://github.com/llm-platform-security/chatgpt-plugin-eval/blob/main/aiagents-interaction.pdf}{AI Agents interaction link}~\cite{aiagents-interaction}.
We conducted this analysis on June 09, 2023.

Similarly, OpenAI prohibits developers from using the words like \textit{plugin}, \textit{ChatGPT}, or \textit{OpenAI} in the plugin name or description and requires that they use \textit{correct grammar} in their functionality descriptions~\cite{openai_plugin_store}.  
However, we found several plugins hosted on the OpenAI's plugin store which do not meet this criteria. 
SEO~\cite{seo_plugin} is one such plugin, which makes spelling and grammatical mistakes in its functionality description.
For example, ``this'' and ``occurrence'' are misspelled as ``ths'' and ``occurance'', respectively.
Another plugins, called edX~\cite{edX_plugin} mentions both \textit{ChatGPT} and \textit{plugin} in its description.

\end{document}